\documentclass[pra,english,preprintnumbers,amsmath,amssymb,nofootinbib,twocolumn,superscriptaddress]{revtex4}
\pdfoutput=1
\usepackage[latin1]{inputenc}
\usepackage{graphicx}
\usepackage{color}
\usepackage{bbm}
\usepackage{amssymb}
\usepackage{amsmath}
\usepackage{tabularx}
\usepackage{dsfont}
\usepackage{multirow}
\usepackage{dcolumn}

\def\0#1#2{\frac{#1}{#2}}

\def\s0#1#2{\mbox{\small{$ \frac{#1}{#2} $}}}

\def\CU{{\mathcal U}}

\newcommand{\beq}{\begin{equation}}
\newcommand{\eeq}{\end{equation}}
\newcommand{\bea}{\begin{eqnarray}}
\newcommand{\eea}{\end{eqnarray}}
\newcommand{\tr}{\mathrm{tr}}

\usepackage{babel}
\makeatother
\begin{document}

\title{Third-order perturbative lattice and complex Langevin analyses of the finite-temperature equation of state of non-relativistic fermions in one dimension}

\author{Andrew C. Loheac}
\affiliation{Department of Physics and Astronomy, University of North Carolina, Chapel Hill, North Carolina 27599, USA}

\author{Joaqu\'{i}n E. Drut}
\affiliation{Department of Physics and Astronomy, University of North Carolina, Chapel Hill, North Carolina 27599, USA}

\begin{abstract}
We analyze the pressure and density equations of state of unpolarized non-relativistic fermions at finite temperature in one spatial dimension.
For attractively interacting regimes, we perform a third-order lattice perturbation theory calculation,
assess its convergence properties by comparing with hybrid Monte Carlo results (there is no sign problem in this regime), and demonstrate agreement
with real Langevin calculations. For repulsive interactions, we present lattice perturbation
theory results as well as complex Langevin calculations, with a modified action to prevent uncontrolled excursions in the complex plane. Although 
perturbation theory is a common tool, our implementation of it is unconventional; we use a Hubbard-Stratonovich transformation to decouple
the system and automate the application of Wick's theorem, thus generating the diagrammatic expansion, including symmetry factors, at any
desired order. We also present an efficient technique to tackle nested Matsubara frequency sums without relying on contour integration, which
is independent of dimension and applies to both relativistic and non-relativistic systems, as well as all energy-independent interactions.
We find exceptional agreement between perturbative and non-perturbative results at weak couplings, and furnish predictions based on complex
Langevin at strong couplings. We additionally present perturbative calculations of up to the fifth-order virial coefficient for repulsive and attractive couplings.
Both the lattice perturbation theory and complex Langevin formalisms can easily be extended to a variety of situations including polarized systems, bosons,
and higher dimension.
\end{abstract}

\maketitle
\section{Introduction}

By far most problems in quantum many-body physics, from quantum chemistry to condensed matter and high-energy areas, suffer from
the so-called sign problem. This problem is an exponential deterioration of the signal-to-noise ratio that appears when estimating
quantum expectation values stochastically (i.e. via some form of quantum Monte Carlo), and is present in nearly all non-relativistic models at
finite polarization and relativistic ones at finite chemical potential (see e.g.~\cite{SignProblemReview1,SignProblemReview2,SignProblemReview3}
for reviews). While sufficient conditions for the avoidance of the sign problem have been under intense study recently~\cite{ZhangEtAl, LiEtAl},
a general solution has been shown to be a formidable challenge~\cite{TroyerWiese}.
Motivated in part by the above, we set out to examine a classic many-body problem normally plagued by a sign problem using both
perturbative and non-perturbative methods, in all cases regulated by a spatial lattice. Specifically, we focus on
a system of non-relativistic fermions in one spatial dimension (1D) with a contact interaction, for which a sign problem is present
when interactions are repulsive.

On one hand, we calculate analytically the first few orders of the lattice perturbative expansion of the pressure, aiming to assess
their reliability towards characterizing thermodynamic equations of state. The results of this calculation provide a posteriori motivation
for our present work: First, the perturbative finite-temperature lattice results and formalism shown here are unconventional, as perturbative calculations are typically
done in the continuum; second, we have found methodological aspects that also appear to have been previously overlooked, which greatly simplify
the calculation on the lattice, and which apply to a very wide range of systems (including relativistic ones); third, we consider 1D as
an initial benchmarking step towards more demanding, higher-dimensional calculations; finally, semi-analytic tools like perturbation theory
can address the shortcomings of conventional quantum Monte Carlo calculations where the sign problem is present, at least in
some regimes.

On the other hand, we extend the hybrid Monte Carlo calculations of Ref.~\cite{EoS1D} and study the repulsive case by applying complex Langevin (CL) techniques.
The latter have garnered considerable attention in the last few years in the area of finite-density lattice QCD~\cite{FiniteDQCD0, FiniteDQCD1,FiniteDQCD2,FiniteDQCD3,FiniteDQCD4,FiniteDQCD5,FiniteDQCD6,FiniteDQCD7,FiniteDQCD8},
but their application to non-relativistic matter is almost nonexistent (see however~\cite{CLYamamoto1,CLYamamoto2}).
We use the CL method with a modified action to prevent uncontrolled excursions into the complex plane, which would otherwise
lead the method to converge to an incorrect result.
While we focus here on a 1D problem, all the methods used extend easily to arbitrary dimensions, as well as to polarized and multi-component
systems. However, a comment is in order that is specific to 1D: As pointed out in Ref.~\cite{BA}, in 1D one may apply the thermodynamic Bethe ansatz in certain
finite-temperature regimes, but it is generally an uncontrolled approximation as it involves solving an infinite tower of non-linear integral equations
(see however Ref.~\cite{KakashviliBolech}).

Our motivation for studying 1D fermions is in part computational convenience, as mentioned above, and in part due to the fact that the general 1D
quantum many-body problem remains an area of great interest for condensed matter, atomic, and high-energy physics. Substantial progress was made
in the last few decades in particular for the 1D Hubbard model (see~\cite{1DHubbardBook}) using numerical methods such as quantum Monte Carlo
(see e.g.~\cite{1DMC1,1DMC2,1DMC3}), density matrix renormalization group (see e.g.~\cite{DMRG1,DMRG2,DMRG3,DMRG4}), and the Bethe ansatz
(see~\cite{BA} and~\cite{BatchelorEtAl,BatchelorEtAlGaudinYang1,BatchelorEtAlGaudinYang2}), as
well as with analytic approaches (such as bosonization~\cite{Bosonization} and beyond-mean-field approaches~\cite{RBD}). Remarkably, there are aspects of this problem that remain elusive
in the sense that they lie beyond the Luttinger liquid paradigm, as pointed out
in Ref.~\cite{RMPGlazman}, which justifies more detailed studies. On the experimental side, ultracold atoms in optical lattices
continue to provide an unparalleled realization of clean, malleable, fermionic and bosonic systems, and thus these systems continue
to shed light on multiple aspects of strongly coupled quantum dynamics (see~\cite{ExpReviewLattices,1DExp,1DSUNExp}).

Below we present the lattice perturbation theory formalism (Sec.~\ref{Sec:Formalism}), which is unconventional in that it uses a Hubbard-Stratonovich (HS) transformation~\cite{HS1,HS2} to facilitate
the automation of generating a diagrammatic expansion. With the diagrams in hand, the evaluation of physical observables proceeds in two steps: first, the Matsubara-frequency sums are done with a technique that allows us to carry them out simultaneously (rather than iteratively, as is the case when using countour integrals,
see e.g. Ref.~\cite{Nieto}); second, the remaining sum over spatial momenta is calculated numerically. Performing the frequency sums analytically is primarily avantageous in that it yields a vastly improved scaling of the computation time, as the total number of nested sums is greatly reduced.
In Sec.~\ref{Sec:CLFormalism} we present a brief overview of the CL method, contrast it with conventional hybrid Monte Carlo, and introduce a modified action to
stabilize the calculation.
In Sec.~\ref{Sec:Results} we show our results for the pressure and density equations of state up to next-to-next-to-next-to (N3LO) leading order in perturbation theory for both
attractive and repulsive couplings, and compare with hybrid Monte Carlo for the case of attractive interactions where no sign problem is present. We
also demonstrate results for a modified-action form of real and complex Langevin calculations for both attractive and repulsive couplings, and compare with
our perturbative results. Using particle-number projection, we also provide the corresponding perturbative
expansion for the first few virial coefficients. In Sec.~\ref{Sec:Conclusions} we summarize and present our conclusions. The appendices contain
further details on the Matsubara sums involved and how to calculate them.

\section{Perturbation Theory Formalism~\label{Sec:Formalism}}
We first present the formalism for our method of computing the perturbative expansion for the grand-canonical partition function of interacting quantum gases
on the lattice.
In a nutshell, the procedure involves analytically evaluating expressions for the diagrammatic expansion and the resulting path integral on a computer
using an object-oriented approach. The final expressions, which contain sums over the complete frequency-momentum basis, are evaluated numerically to
determine the pressure equation of state, or other physical observable, such as the density equation of state or projected virial coefficients.

\subsection{Path integral form of the partition function}
The starting point of our lattice perturbative expansion of the equation of state is the grand-canonical partition function,
\beq
\mathcal Z = \tr \left[ e^{-\beta(\hat H - \mu \hat N)} \right],
\eeq
where, as usual, $\hat H$ is the Hamiltonian, $\hat N$ is the particle number operator, $\beta$ the inverse temperature, and $\mu$
the chemical potential; the trace is over all the multiparticle states in the Fock space. We will assume that
\beq
\hat H = \hat T + \hat V,
\eeq
where $\hat T$ is the kinetic energy operator, and $\hat V$ is the potential energy operator.
Rather than expanding directly in powers of $\hat V$, as is conventional in perturbation theory,
we will introduce a HS transformation. We first discretize the imaginary time direction as $\beta = \tau N_\tau$, and apply a
 first-order Suzuki-Trotter factorization such that
\beq
e^{-\tau \hat H} \simeq e^{-\tau \hat T/2} e^{-\tau \hat V} e^{-\tau \hat T/2},
\eeq
and next perform the HS transformation,
\beq
e^{-\tau \hat V} = \int \mathcal D \sigma e^{-\tau \hat W[\sigma]},
\eeq
where $\hat W[\sigma]$ is a one-body operator representing an external potential set by the auxiliary field $\sigma$,
and the field integral is over all possible configurations of that field.
The specific form of $\hat W[\sigma]$ depends on the choice of the HS transformation; in the case of fermions it can be either discrete or
continuous, and in the latter case it can be compact or non-compact~\cite{MCReview}. For the purposes of this work, the specific form of the HS transform
is immaterial, as we will undo the transformation order-by-order in the perturbative expansion.

Naturally, the use of the HS transform is much more common in Monte Carlo approaches than in perturbative ones, and
it is the natural path to mean-field theory. Here, however, we will use
it as a device to recover the results of Wick's theorem but bypassing the operator algebra completely, as we show below. In addition,
this approach has the advantage of facilitating the use of many-body forces, as it is relatively simple to write down HS transforms for them and
the steps after the transform are quite mechanical (which is the main reason for our choice).

Collecting all the field integrals and performing the trace over the resulting product of exponentials of one-body operators
yields a path-integral form of the grand-canonical partition function $\mathcal{Z}$,
\beq
\mathcal{Z} = \int \mathcal{D}\sigma \, {\det}^2 M[\sigma],
\eeq
where in the above form we have assumed that the system contains two identical species and the matrix
$M$ of dimension $N_\tau \times V$ by $N_\tau \times V$ (where $N_\tau$ is the temporal volume and $V$ is the spatial volume) encodes all relevant system dynamics, such that
\beq
M[\sigma] \equiv
\begin{pmatrix}
1 & 0 & \cdots & \CU_{N_\tau}[\sigma] \\
-\CU_1[\sigma] & 1 & \cdots & 0 \\
0 & -\CU_2[\sigma] & \ddots & 0 \\
\vdots & \vdots & \ddots & \vdots \\
0 & \cdots & -\CU_{N_\tau - 1}[\sigma] & 1
\end{pmatrix} .
\eeq
For the case of contact interactions, such as in the Gaudin-Yang model, the matrix $\CU[\sigma]$ can be written as
\beq
\left[ \CU_j[\sigma] \right]^{}_{{\bf x},{\bf x}^\prime} \equiv \left[ e^{-\tau({\bf p}^2/2m-\mu)} \right]^{}_{{\bf x},{\bf x}^\prime}\left[1 + A\sin \sigma({\bf x}^\prime,t)\right]
\eeq
where ${\bf x}$ and ${\bf x}^\prime$ are spatial indices, $t$ is a temporal index, and where $A \equiv \sqrt{2(e^{\tau g}-1)}$, $g$ is the zero-range coupling, and $\tau$ is the temporal lattice spacing (see e.g.~\cite{EoS1D}).
We have chosen a realization of the HS transformation in which the continuous field $\sigma$ takes values between $-\pi$ and $\pi$
at each point in spacetime, such that
\beq
\int \mathcal D \sigma = \prod_{{\bf x},\tau} \int_{-\pi}^{\pi} \frac{d\sigma({\bf x},\tau)}{2\pi}.
\eeq
This particular kind of HS transformation was first explored in Refs.~\cite{HSLee1,HSLee2}.

Note that in the case of non-identical species, such as in polarized or mass imbalanced systems of two flavors, two different determinants are present in the partition function, i.e.
\beq
\mathcal{Z} = \int \mathcal{D}\sigma \, \det M_{\uparrow}[\sigma]\det M_{\downarrow}[\sigma].
\eeq

At this point, we have eliminated from the problem all the quantum operators, and this is one of the main advantages
of the method. Moreover, we have accomplished this by ``integrating out'' the fermionic degrees of freedom,
which is an unorthodox route to perturbation theory. We will see, however, that this is a useful way to proceed in the sense that
it mechanically generates the correct answers, and therefore it is amenable to automation.

\subsection{Expanding the fermion determinant}

To obtain the perturbative expansion of the grand-canonical partition function on the lattice, we expand the determinant of the matrix $M$ in powers of the
dimensionless parameter $A$ defined above. To this end, first note that
\beq
M = M_0 + A \mathcal{T} \mathcal{S}[\sigma]
\eeq
where $M_0$ is the non-interacting counterpart, and
\beq
\mathcal{T} \equiv \begin{pmatrix}
0 & 0 & \cdots & T \\
-T & 0 & \cdots & 0 \\
0 & -T & \ddots & 0 \\
\vdots & \vdots & \ddots & \vdots \\
0 & \cdots & -T & 0
\end{pmatrix},
\eeq
with $\left[T \right]_{{\bf x},{\bf x}^\prime} \equiv \left [ e^{-\tau (\mathbf{p}^2/2 - \mu)} \right ]_{{\bf x},{\bf x}^\prime}$,
(where we have taken $\hbar = k_B = m = 1$), and
\beq
\left[ \mathcal{S}[\sigma]  \right]_{{\bf x},t;{\bf x}^\prime,t'} = \delta_{t,t'}\delta_{{\bf x},{\bf x}^\prime} \sin\sigma({\bf x},t).
\eeq
Thus, we further obtain that $M =  M_0 ( \openone + A K\mathcal{S}[\sigma])$,
where $K \equiv M_0^{-1} \mathcal{T}$ will play the role of the free propagator, as we explain below, and of course
\beq
\det M = \det M_0 \det ( \openone + A K\mathcal{S}[\sigma]).
\eeq
At this point, we set aside the non-interacting factor and make use of the identity $\det = \exp \tr \ln$ to formally expand the logarithm
in powers of $A$, such that
\bea
\det ( \openone + A K\mathcal{S}[\sigma])
&=&
\exp\left[ \sum_{k=1}^{\infty} A^k E_k\right],
\eea
where
\beq
\label{Eq:Ek}
E_k = \frac{(-1)^{k+1}}{k}\tr \left[\left(K\mathcal{S}[\sigma]\right)^k\right].
\eeq
Note that each power of the fermion determinant translates simply as a prefactor in the definition of $E_k$. Thus, the expansion coefficients for an unpolarized system of $N_f$ species are simply $N_f E_k$. We further write the determinant as
\bea
\exp\left[ \sum_{k=1}^{\infty} A^k E_k\right]
&=&
\prod_{k=1}^{\infty}\exp\left[ A^k E_k\right]\\
&=&
\prod_{k=1}^{\infty}\sum_{\left\{ i_k \right\} = 0}^{\infty} \frac{(A^k E_k)^{i_k}}{i_k!}.
\eea
Eq.~(\ref{Eq:Ek}) begins to expose the powers of $A$ in the perturbative expansion. The task ahead is to isolate those powers and carry out the path integral exactly. Using these expressions, by fully expanding and combining terms of identical powers of $A$, it is straightforward to see that
\begin{multline}
\det ( \openone + A K\mathcal{S}[\sigma]) = 1 + A E_1 + A^2 \left(\frac{E_1^2}{2!} + E_2 \right) \\
+ A^3 \left ( \frac{E_1^3}{3!} + E_1 E_2+ E_3  \right) + \cdots .
\end{multline}

The above expansion corresponds to one of the spin degrees of freedom and therefore it is to be combined with
a corresponding expansion for the other spin, which could be done for polarized (different $\mu$ for each species)
or unpolarized (same $\mu$) cases. Here we focus on the latter with a two-component system, which simply means that we are expanding
the square of the determinant. As mentioned above, this can be accomplished easily by setting $E_k \to N_f E_k$
for the general case of $N_f$ flavors.

\subsection{Recovering Wick's theorem by calculating the path integral exactly at each order}

Once the determinants are expanded to the desired power of $A$, which is easy to automate, one obtains products of powers of the
various $E_k$ defined in Eq.~(\ref{Eq:Ek}). The path integral of each of these products must be evaluated to obtain the expansion
of the partition function. For instance, at second order, one of the terms is
\bea
\int \mathcal D \sigma E_2[\sigma] &=& -\frac{1}{2} \int \mathcal D \sigma\; \tr \left[\left(K\mathcal{S}[\sigma]\right)^2\right]\nonumber \\
&=& -\frac{1}{2}  K_{ij}K_{km}\int \mathcal D \sigma\; \mathcal{S}[\sigma]_{jk} \mathcal{S}[\sigma]_{mi},
\eea
where all the (collective, spacetime) indices $i,j,k,m$ on the right-hand side are assumed to be summed over.

Writing out the indices explicitly at each order, we see that the main problem is computing path integrals of the form
\begin{multline}
I_m(b_1,b_2,\dots,b_m) = \\   \int \mathcal D \sigma \sin(\sigma[b_1])\sin(\sigma[b_2]) \dots \sin(\sigma[b_m]) \nonumber,
\end{multline}
where each $b_k$ represents a spacetime coordinate.
These integrals vanish if $m$ is odd, but otherwise the result is generally finite and positive. For instance, for $m=2$,
\bea
I_2(b_1,b_2) &=& \frac{1}{2} \delta_{b_{1},b_{2}},
\eea
and for $m=4$,
\bea
\label{Eq:I4Tensor}
I_4(b_1,b_2,b_3,b_4) =&&
\!\!\!\!\!\!\!\!\!\! \frac{3}{8}\delta_{b_{1},b_{2}} \delta_{b_{1},b_{3}} \delta_{b_{1},b_{4}} \nonumber \\
+\frac{1}{4}\!\!\!\!\!\!\!\!\!\!\!&&
\left[
\delta_{b_{1},b_{2}}\delta_{b_{3},b_{4}}(1-\delta_{b_{2},b_{3}}) \right . \nonumber\\
&+&\delta_{b_{1},b_{3}}\delta_{b_{2},b_{4}}(1-\delta_{b_{3},b_{2}}) \nonumber\\
&+&\left . \delta_{b_{1},b_{4}}\delta_{b_{2},b_{3}}(1-\delta_{b_{4},b_{2}})
\right],
\eea
where $3/8$ is the result for the case of all four coordinates coinciding, i.e.
\beq
\int_{-\pi}^{\pi} \frac{d \sigma}{2\pi} \sin^4\sigma = \frac{3}{8}.
\eeq

It is the tensor expressions like Eq.~(\ref{Eq:I4Tensor}) that automatically implement Wick's theorem when contracting
with the various $K$ propagators. As with the Taylor expansion of the determinant, the calculation of the $I_m$ needed for
each of the terms at a given order was also automated, as was the subsequent index contraction with the propagators.
The result of that process is that not only the diagrams themselves, but also all the symmetry factors are generated automatically,
minimizing the amount of manual bookkeeping. Naturally, the topology of the diagrams enters through the $I_m$ tensors, which encode
the multiple ways in which the corresponding path integral can give a non-vanishing result.

The complexity of the expression for $I_m$ grows with each order in the perturbative expansion, and causes the number
of terms in the expansion of $\mathcal{Z}$ (i.e. after contracting with the $K$'s) to grow very quickly. Although the next-to-leading
(NLO) contribution can easily be verified by hand, the next-to-next-to-next-to-leading (N3LO) (i.e. $A^6$) order produces (naively)
on the order of $10^3$ terms, all of which must be simplified to obtain the final results. Moreover, in the case of polarized systems, multiple
products of the determinant must be expanded. This scaling underscores why the method we proposed here is well
suited for automation but it is otherwise not ideal for manual calculation, especially if a high-order, finite-temperature expansion is
the goal.

\subsection{Transforming to frequency-momentum space on the lattice}

Naturally, the matrix $K$ can be diagonalized in the momentum basis
with a discrete Fourier transform, such that
\beq
[K]_{ab} = \sum_{q} U^{\dagger}_{aq}[D_0]_{q}U_{qb} \label{Eq:FourierIdentity}
\eeq
where $a,b$ are collective spacetime indices of the form $(t,x)$, and $q = (\omega,k)$ is a collective frequency-momentum index,
with $\omega = (2 n_\omega + 1)/N_\tau$ being a fermionic Matsubara frequency ($n_\omega=1,\dots,N_\tau$), and $\mathbf{k}=2 \mathbf{n_k}\pi/N_x$
($\mathbf{n}_{\mathbf{k},i} = 1,\dots,N_x$)
a $d$-dimensional spatial wavevector.
The free propagator $D_0$ is then
\beq
[D_0]_{q} \equiv \frac{1}{1-\exp[i\omega + \tau(\mathbf{k}^2/2-\mu)]},
\eeq
while the Fourier transform matrices take the form
\beq
U_{aq} = \frac{1}{\sqrt{N_x^d N_\tau}} \exp[i(\omega t - \mathbf{k} \cdot \mathbf{x})].
\eeq
By computing the path integral over the auxiliary field $\sigma$ for all the terms in the expansion of the determinant
at a particular order, we are left with an object of the generic schematic form
\beq
S = \sum [K]_{a^{}_1 a^{}_2}[K]_{a^{}_3 a^{}_4}\dots[K]_{a^{}_{2m-1}a^{}_{2m}}I_{a^{}_1\dots a^{}_{2m}}
\eeq
where the sum is taken over all free spacetime indices, and the object $I_{a_1\dots a_{2m}}$
results from the path integral over the interaction terms $\sin(\sigma_i)$.

Upon inserting the Fourier representation of the propagator [c.f. Eq. (\ref{Eq:FourierIdentity})], we obtain a form
that is described by a collection of indices in the frequency-momentum basis:
\beq
S = \sum [D_0]_{q_1} [D_0]_{q_2} \dots [D_0]_{q_m} \tilde{I}_{q_1,\dots,q_{m}}
\eeq
where $\tilde{I}_{q_1,\dots,q_m}$ results from appropriately contracting the various Fourier tensors $U$ and $U^\dagger$
with $I_{a^{}_1\dots a^{}_{2m}}$. The $\tilde I$ tensors represent momentum conservation laws for each
specific term.

The advantage of going to frequency-momentum space is that $S$ can be obtained by performing
$m$ frequency-momentum sums instead of the $2m$ spacetime sums. This optimization, however, is not enough;
it is crucial to carry out the frequency sums analytically in order to have a numerically manageable expression at the end.
We turn to those sums next.

\subsection{Computing finite Matsubara frequency sums analytically: two tricks}
\label{Section:FrequencySums}

In the evaluation of expressions at a given order $n$, we are faced with nested frequency
sums that schematically look like the expression
\beq
\sum_{q_1,q_2,\dots,q_m} [D_0]_{q_1} [D_0]_{q_2} \dots [D_0]_{q_m} \delta(\{q_k\}),
\eeq
where we have left out sums over momenta, which are to be carried out afterwards, and the delta
factor represents an energy-momentum conservation law that is derived from each diagram's topology. Note
that, here and below, we will use the delta notation to represent the discrete Kronecker symbol rather than
the Dirac symbol (we have no need for Dirac deltas here, as all our expressions are discrete).
In this section, we show how we performed the frequency sums in a way that does not use complex contour
integration and, moreover, allows us to treat all the sums simultaneously.

We begin with an example. The simplest case is that of a single factor, which may seem trivial but is nevertheless instructive:
\beq
S_1 = \sum_{n=1}^{N_\tau} [D_0]_{n} = \sum_{n=1}^{N_\tau} \frac{1}{1 - Q \exp{(i\omega_n)}},
\eeq
where we have encoded all the momentum and chemical potential dependence in the factor $Q$,
and where $\omega_n = (2n + 1)\pi / N_\tau$. In this case we expand the expression inside the sum
as a geometric series:
\bea
S_1 &=&
\sum_{n=1}^{N_\tau} \sum_{k=0}^{\infty} Q^k e^{ i\omega_n k}
=
\sum_{k=0}^{\infty} Q^k \sum_{n=1}^{N_\tau}  e^{ i\omega_n k} \\
&=&
\sum_{m=0}^{\infty} \sum_{k=0}^{\infty} Q^k \delta(k - m N_\tau)(-1)^{m} \\
&=&
\sum_{m=0}^{\infty} Q^{m N_\tau} (-1)^{m} =
\frac{N_\tau}{1 + Q^{N_\tau}}, \label{Eq:S1_0}
\eea
where we used the fact that $k \geq 0$ and
\beq
\sum_{n=1}^{N_\tau}  e^{ i 2\pi n k /N_\tau} = N_\tau \sum_{m=-\infty}^{\infty} \delta(k - m N_\tau).
\eeq

The $S_1$ sum is useful per se, but also because high-order calculations need the more general sums,
such as
\bea
S_1^{(k)} = \sum_{n=1}^{N_\tau} \left(\frac{1}{1 - Q \exp{(i\omega_n)}} \right)^{k+1},
\eea
such that $S_1^{(0)} = S_1$. For general $k$, this is easy to compute by introducing a new parameter $\lambda$ via
\bea
\label{Eq:S1LambdaInsertion}
S_1(\lambda) &\equiv&
\sum_{n=1}^{N_\tau} \frac{1}{\lambda - Q \exp{(i\omega_n)}} = \frac{1}{\lambda} \frac{N_\tau}{1 + (Q/\lambda)^{N_\tau}},
\eea
and differentiate with respect to $\lambda$ as needed,
\beq
S_1^{(k)} = \frac{(-1)^k}{k!}\left. \frac{d^{k}S_1(\lambda)}{d\lambda^k} \right |_{\lambda=1}.
\eeq
Clearly, $S_1(\lambda)$ is a generating function for frequency sums of higher order. In particular, for instance,
\beq
\label{Eq:S1_1}
S_1^{(1)} = \frac{[1 - Q^{N_\tau} (N_\tau - 1)] N_\tau} {(1 + Q^{N_\tau})^2}.
\eeq
%


The method of expanding the numerator as a power series applies as well when multiple sums are present, as
we show in Appendix~\ref{AppendixBeachBall} with an example that corresponds to the ``beach-ball'' diagram
of Fig.~\ref{Fig:PTDiagramsN2LO}(b).

The general procedure of the method is as follows. First, take a step backward and use the Fourier-sum representation of all
the frequency Kronecker deltas, i.e.
\beq
\delta[f(\omega_1, \omega_2, \dots)] = \frac{1}{N_\tau}\sum_{p=0}^{N_\tau-1} e^{i p f(\omega_1, \omega_2, \dots)},
\eeq
in combination with the geometric series expression of the denominators. This first step is similar to that of techniques
used in the continuum, where an integral representation of a Dirac delta function is utilized. Beyond this point, however,
the calculations differ considerably.

The second step is to sum over each fermionic frequency
$\omega_n$, which yields as many delta functions as denominators, with their corresponding ``boundary''
sums and factors of $(-1)^m$, i.e. use
\beq
\sum_{n=1}^{N_\tau} e^{i \omega_n y} = N_\tau \sum_{m=-\infty}^{\infty} \delta( y - m N_\tau ) (-1)^m.
\eeq
Here it is important to keep the terms for all $m$, as $y$ will generally vary over a semi-infinite range due
to the geometric series expansion.

Third, sum over the geometric sum index to saturate the delta functions generated in the previous step. To this end, it is useful to
extend the geometric sum index to $-\infty$, such that Heaviside step functions should be inserted accordingly.

Fourth, implement the constraints over the boundary sums and evaluate such sums, which lead to the expected $(1 + Q_k^{N_\tau})^{-1}$ factors.
Finally, sum over the Fourier index of the energy-conserving delta functions.
As is clear from the above, the operations to be performed are rather elementary and do not involve complex analysis, only
a small amount of bookkeeping. It is the order of the operations that is crucial for the simplicity of the method.

\section{Complex Langevin Approach \label{Sec:CLFormalism}}

The second method we present in this work involves the use of CL dynamics in the context of the lattice quantum Monte Carlo technique,
which is, of course, a nonperturbative approach.
Recent developments have shed light on several aspects of the CL method, which had previously hindered progress but which are now
starting to be better understood and in some cases even resolved. In this section we discuss our implementation of CL dynamics, including
a kind of dynamical stabilization that is reminiscent of the one put forward in Refs.~\cite{DynamicStabilisation1,DynamicStabilisation2}.

The starting point is once again the partition function
\beq
\mathcal{Z} = \int \mathcal{D}\sigma \, {\det}^2 M[\sigma],
\eeq
where one normally identifies $P[\sigma] = {\det}^2 M[\sigma]$ as the (unnormalized) probability measure to be used in a Metropolis-based
Monte Carlo calculation. One may define an effective action $S$ via
\beq
P[\sigma] = \exp(-S[\sigma]).
\eeq

Observables, at least simple ones, are then shown to take the form
\beq
\langle \mathcal O \rangle = \frac{1}{\mathcal Z} \int \mathcal D \sigma \, e^{-S[\sigma]} \mathcal O[\sigma],
\eeq
such that the expectation value can be determined by sampling the auxiliary field $\sigma$ according to $P[\sigma]$. Two well known ways of
carrying out that sampling, without the presence of a sign problem, are the hybrid Monte Carlo algorithm (HMC)~(see Refs.~\cite{HMC1,HMC2})
and the real Langevin method (RL)~(also known simply as stochastic quantization; see Refs.~\cite{RL1,RL2,RL3}).

In HMC, one defines an auxiliary field variable $\pi$ conjugate to the HS field $\sigma$ along
with global (molecular-dynamics type) equations of motion in a fictitious phase-space time $t$ such that
\bea
\dot \sigma &=& \pi, \\
\dot \pi &=& -\frac{\delta S[\sigma]}{\delta \sigma},
\eea
where the right-hand side of the last equation naturally represents a molecular dynamics force. These equations are usually integrated via the
leapfrog method (or more sophisticated variants) to ensure reversibility, which in turn is essential for maintaining detailed balance. The global
updates that this method allows are essential for lattice QCD calculations. The new field obtained at the end of the trajectory is
accepted or rejected using the criteria of a Metropolis step.

In RL, on the other hand, there is no Metropolis step and the equations of motion are different:
\bea
\dot \sigma = -\frac{\delta S[\sigma]}{\delta \sigma} + \eta,
\eea
where we note that there is no auxiliary momentum field $\pi$ but a $t$-dependent noise field $\eta$ appears instead.
The latter satisfies $\langle \eta(x,\tau) \rangle = 0$ and $\langle \eta(x,\tau) \eta(x',\tau') \rangle = 2 \delta_{x,x'}\delta_{\tau,\tau'}$.

The (conventional) mathematical underpinnings of HMC and RL depend on $P[\sigma]$ being positive semi-definite.
As is well known, this property generally fails to hold e.g. for repulsive interactions, polarized systems, etc. as mentioned above, and the calculation
is then said to have a sign problem (or more generally a complex phase problem). In the case of HMC, this means that the Metropolis step is simply no
longer well-defined and thus the algorithm is no longer available. For RL, on the other hand, a generalization is possible, namely CL, as first noted in Ref.~\cite{Parisi1983}. In CL, one complexifies
the HS field $\sigma$ into
\beq
\sigma = \sigma_R + i \sigma_I,
\eeq
where $\sigma_R$ and $\sigma_I$ are both real fields. The CL equations of motion are given by
\bea
\delta \sigma_R &=& -\text{Re}\left[\frac{\delta S[\sigma]}{\delta \sigma}\right] \delta t + \eta \sqrt{dt}, \\
\delta \sigma_I &=& -\text{Im}\left[\frac{\delta S[\sigma]}{\delta \sigma}\right] \delta t,
\eea
where now $S[\sigma]$ is to be understood as a complex function of the complex variable $\sigma$.
Note that, when the action is real, the imaginary part of the force is zero and then CL reduces to RL.

Under certain conditions, which have lately received much attention
(see~\cite{CLSufficientConditions00,CLSufficientConditions01,CLSufficientConditions02,CLSufficientConditions03}),
the CL method can be guaranteed to converge to the right answer.
In those cases, expectation values $\langle \mathcal O \rangle$ are correctly obtained by averaging over the real part of
$\mathcal O[\sigma]$ with complex fields $\sigma$ sampled throughout the CL dynamics evolution.

In the process of making CL a viable solution to the sign problem in lattice QCD, two crucial challenges were identified: the appearance
of numerical instabilities in the CL evolution and the uncontrolled excursions of $\sigma$ into the complex plane due to singularities in $P[\sigma]$.
The former was largely resolved by implementing adaptive time-step solvers~\cite{CLAdaptiveSolvers}; the latter, on the other hand, is currently under
investigation and a few approaches have been proposed (see e.g.~\cite{RCL1, RCL2}).
In the calculations presented here, those excursions are highly problematic because the dependence of $P[\sigma]$ on $\sigma_I$ is through
hyperbolic functions; indeed, the HS transformation we use depends on $\sin \sigma$, and
\beq
\sin \sigma = \sin (\sigma_R) \cosh(\sigma_I) + i \cos (\sigma_R) \sinh(\sigma_I).
\eeq
Thus, a growing (positive or negative) $\sigma_I$ effectively increases the coupling at an exponential rate, which can completely stall the calculation
or result in a converged but wrong answer (as we have observed in our tests). This exponential growth is similar to the problem found in gauge theories,
as the complexified link variables representing the gauge field become unbounded in the same fashion in those theories.

\subsection{Modified action}

To overcome the problem of large excursions in the complex plane in the CL algorithm, we modified the action in a way
reminiscent of the dynamical stabilization approach of Ref.~\cite{DynamicStabilisation1,DynamicStabilisation2}. In the latter, 
a new term was added to the CL dynamics which vanishes in the continuum limit and renders the calculation stable. We 
propose here to modify the CL equations by adding a regulating term controlled by a parameter $\xi$, such that the new equations, 
in their discretized form, are
\bea
\delta \sigma_R &=& -\text{Re}\left[\frac{\delta S[\sigma]}{\delta \sigma}\right] \delta t  - 2 \xi \sigma_R \delta t + \eta \sqrt{\delta t}, \\
\delta \sigma_I &=& -\text{Im}\left[\frac{\delta S[\sigma]}{\delta \sigma}\right] \delta t - 2 \xi \sigma_I \delta t.
\eea
This change amounts to modifying the action by
\beq
S[\sigma] \to S[\sigma] + \xi \sigma^2. \label{Eq:CLModifiedAction}
\eeq
The rationale for adding such a regulating term was originally based on
our understanding of ${\delta S[\sigma]}/{\delta \sigma}$ as a molecular dynamics force in the context of the HMC algorithm (although it is
typically referred to as the ``drift'' in the context of the CL method). In the HMC sense, the new term in the action can be understood as a harmonic oscillator
trapping potential, i.e. a restoring force that prevents the field from running away. However, HMC does not apply when $\sigma$ is complex. A
more appropriate interpretation is obtained by keeping only those new terms in the CL equations and neglecting the rest, which results in the
decoupled form
\bea
\delta \sigma_R &=& - 2 \xi \sigma_R \delta t, \\
\delta \sigma_I &=& - 2 \xi \sigma_I \delta t,
\eea
whose solution is a decaying exponential (assuming $\xi > 0$) for both $\sigma_R$ and $\sigma_I$. Thus, this new ad hoc term
represents a damping force.

\begin{figure}[]
	\centering
	\includegraphics[width=\columnwidth]{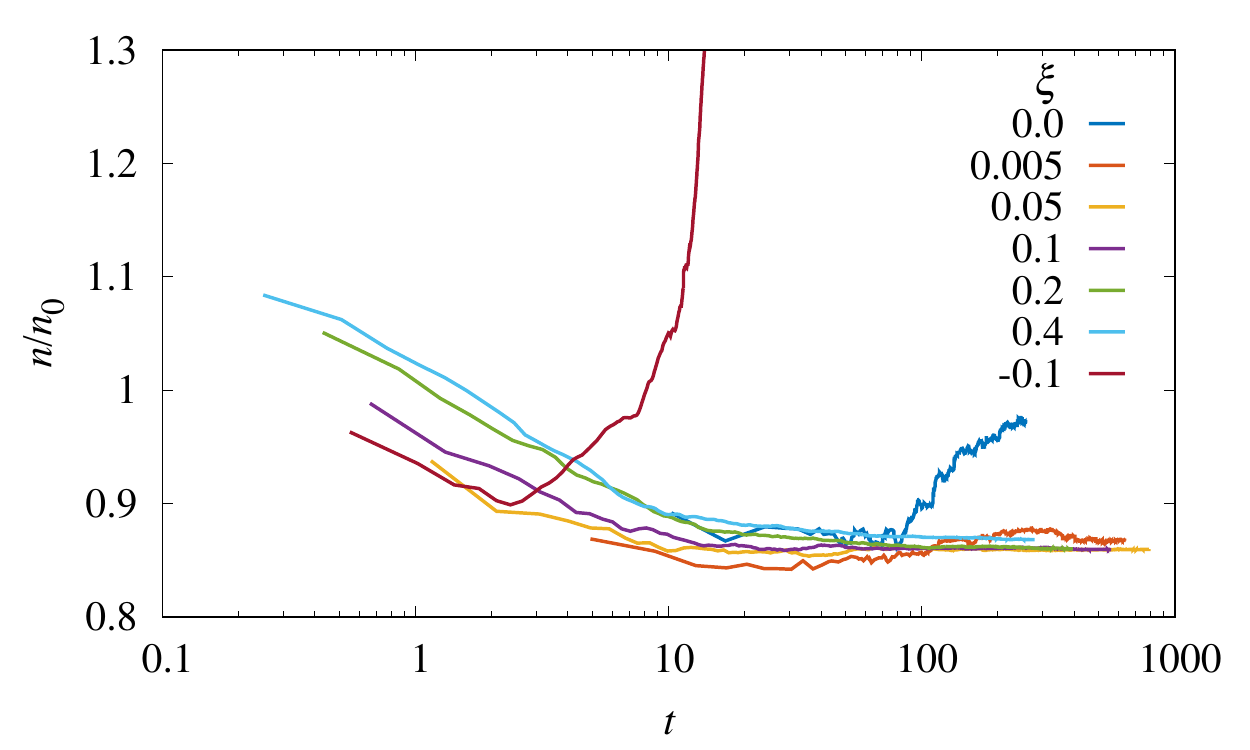}
	\caption{(Color online) The normalized density $n/n_0$, where $n_0$ is the non-interacting result,
	for $\lambda=-1.0$ and $\beta\mu = 1.6$, as a function of the Langevin time $t$ for
	several values for the regulating parameter $\xi$ [see Eq. (\ref{Eq:CLModifiedAction})].
	For a choice of $\xi = 0$, where the regulating term is removed, CL tends toward an incorrect value for the density. When $\xi \simeq 0.1$,
	the additional term provides a restoring force and the stochastic process converges to a different value consistent with perturbation theory.
	On the other hand, for cases
	where $\xi < 0$, the solution diverges, as expected. Each plotted line corresponds to a fixed count of $10^5$ iterations of performing one integration
  step of the adaptive step $\delta t$; as such, the length of the line gives an indication as to the computational demand to reach time $t$
  for a given $\xi$.}
	\label{Fig:XiComparison}
\end{figure}

\begin{figure}[]
	\centering
	\includegraphics[width=\columnwidth]{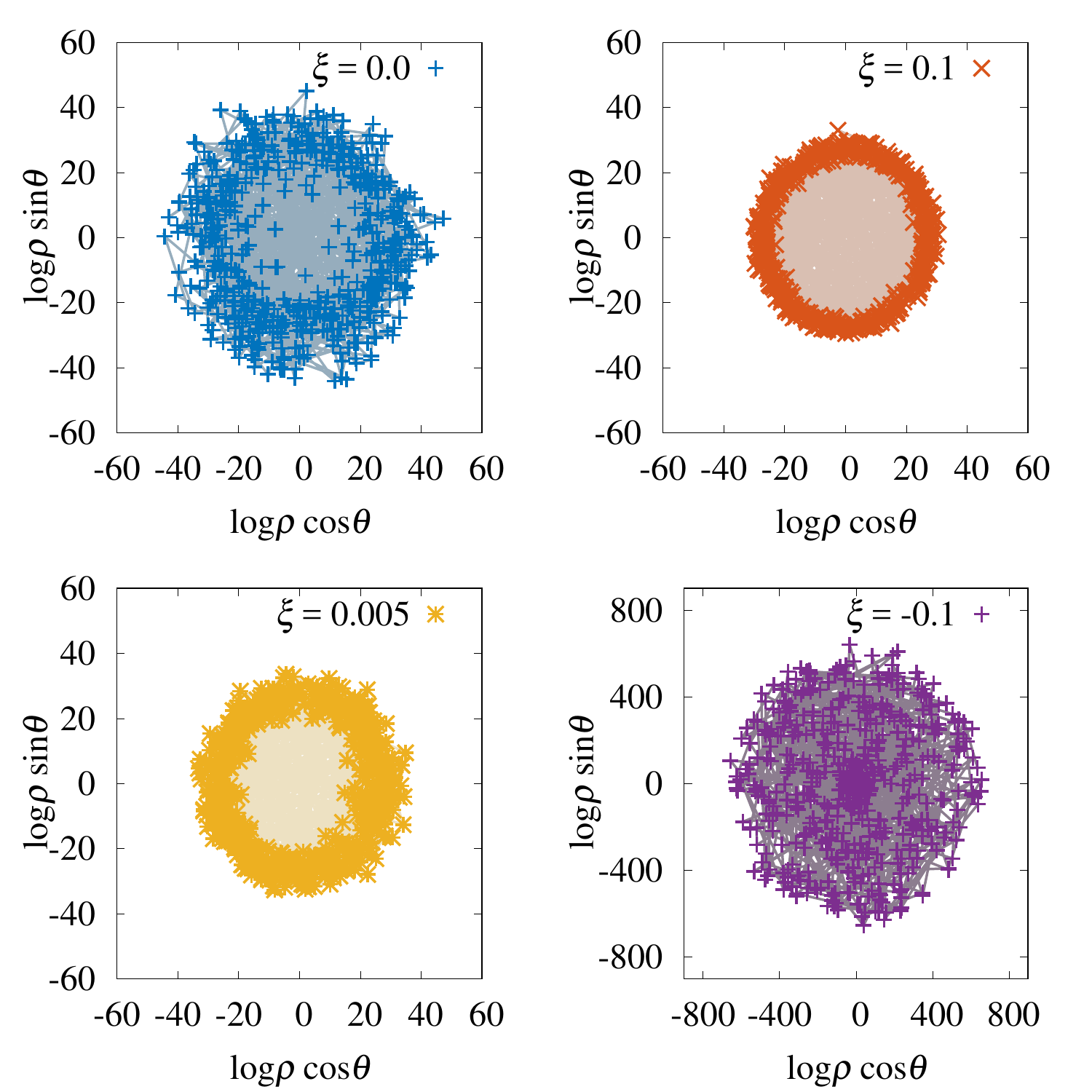}
	\caption{(Color online) Plot of the complex quantity $e^{-S}$ in terms of its magnitude $\rho$ and phase $\theta$ such that $e^{-S} = \rho e^{i\theta}$,
	for a CL calculation at $\lambda=-1.0$ and $\beta\mu = 1.6$.
	Data points are plotted as $\log (\rho) \cos(\theta)$ and $\log (\rho) \sin(\theta)$ as a parametric functions of the Langevin time $t$.
	Plots are displayed for four values of the parameter $\xi$. Note that for $\xi = 0$, the solution does not converge, but does converge
	for $\xi = 0.005$ and $0.1$. For the case where $\xi = -0.1$, the result for the density $n/n_0$ rapidly diverges, as expected
	[note change in scale for $y$ axis and see Fig. \ref{Fig:XiComparison}].
	Data points show the locations where samples were taken along the CL trajectory; the shaded areas result from straight lines joining the data points.}
	\label{Fig:XiCLPhase}
\end{figure}

Naturally, the proposed modification introduces a systematic effect that needs to be studied for each quantity of interest, i.e.
it is crucial to understand the $\xi$ dependence of the output. The results presented below correspond to $\xi=0.1$. To gain specific insight into
the variations of the density with $\xi$, we show in Fig.~\ref{Fig:XiComparison} a plot of the running average of the density as a function of the
Langevin time $t$ for several values of $\xi$ in the neighborhood of $0$. As evident in that figure, there is a sizable window of small values of $\xi$
where CL converges. Additionally, we show in Fig.~\ref{Fig:XiCLPhase} a plot of the phase and
magnitude of $e^{-S}$, where $S$ is the complex action, for different values of $\xi$. The effect of the $\xi$ term on the CL dynamics can be clearly
seen in that figure (note the change of scale for the $\xi < 0$ plot, where the results are expected to diverge).

In the next section we present our results. Whenever RL or CL results are shown, we have kept $\xi=0.1$. The agreement of RL with HMC
(where the regulating term should not be needed) and the weak dependence of the density on $\xi$ (at least within a window close to $\xi=0$)
support the idea that the damping term has negligible impact on the results within the precision studied here.

\section{Results \label{Sec:Results}}

In this section we show the results of implementing the above two formalisms and methods to the case of unpolarized, spin-$1/2$ fermions in one spatial
dimension. The specific Hamiltonian we explore is the Gaudin-Yang form~\cite{GaudinYang1,GaudinYang2} where $\hat H = \hat T + \hat V$ with
\beq
\hat T = \int dx \;
\sum_{s=\uparrow,\downarrow}\! \hat \psi_s^\dagger({x})\!\left(\!-\frac{\hbar^2}{2m} \frac{d^2}{dx^2}\!\right)\! \hat \psi_s^{}({x}),
\eeq
and
\beq
\hat V = - g \int dx \; \hat n^{}_{\uparrow}({x}) \hat n^{}_{\downarrow}({x}),
\eeq
where $\hat \psi_s^{\dagger}, \hat \psi_s^{}$ are the creation and annihilation operators in coordinate space for particles of
spin $s$, and $\hat n^{}_{s} = \hat \psi_s^{\dagger}\hat \psi_s^{}$ are the corresponding density operators.
There are only two dimensionless parameters characterizing this problem at finite temperature: the fugacity $z = e^{\beta \mu}$
and the coupling $\lambda = \sqrt{\beta} g$.

In all of our results below, perturbative as well as non-perturbative, the system was placed on a
lattices of spatial size $N_x = 60, 80, 100$ and temporal size $N_\tau = 160$. The spatial lattice spacing was set to $1$, thus setting the length
scale for the problem, and the temporal lattice spacing was chosen as $\tau = 0.05$ (in units of the spatial spacing).
Previous studies show that, with those parameters (particularly with $N_x \geq 60$), the equations of state of this system are
within $\leq 5\%$ of the continuum limit result for the values of $\lambda$ studied here.

\subsection{Analytic expressions for the perturbative expansion}
Although the results shown here correspond to unpolarized fermions with two internal degrees of freedom (e.g., spin-1/2 particles), the perturbative and complex Langevin
formalisms can easily be extended to greater degrees of freedom (or, ``flavors") by introducing additional determinants in the integrand of the path integral. Additionally,
although all observables computed here correspond to the one-dimensional system, the perturbative expansion itself is independent of the spatial dimension; one must
simply modify the definition of the wavevector $\mathbf{k}$ to extend the system to higher dimension.

Among the main results of this work are the final expressions for the perturbative contributions to the partition function $\mathcal{Z}$ at each order $n$ up to
next-to-next-to-next-to-leading order (N3LO), which are shown in Table~\ref{PTContributionTable} (explained in detail below). Each order $n$ is to be
accompanied by a factor of $A^{2n}$, but all the information required to reproduce the numerical results in this work are otherwise presented in that table and in the appendices.

\begin{figure}[]
	\centering
	\includegraphics[width=110pt]{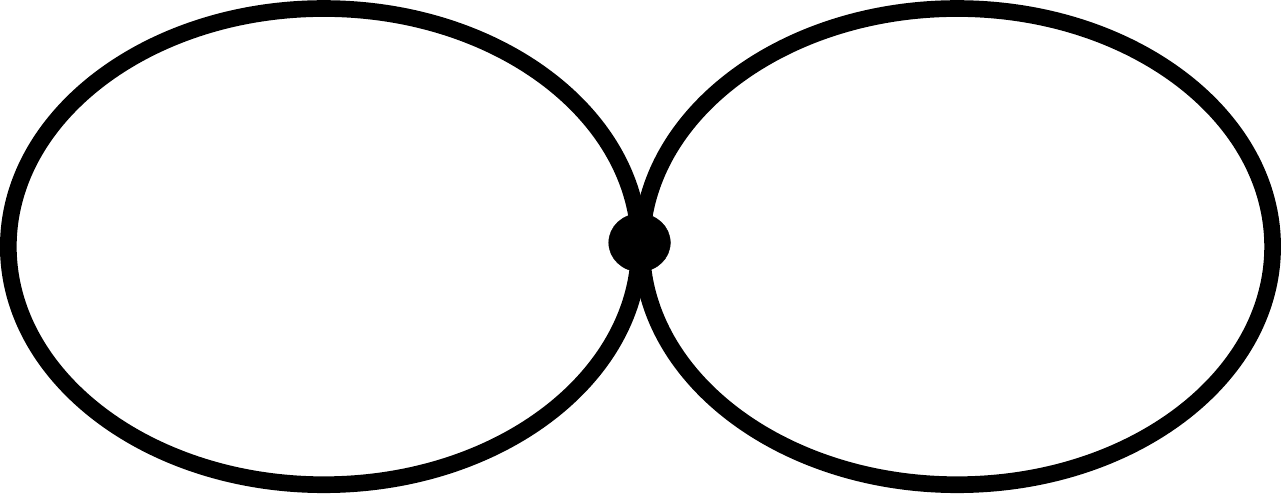}
	\caption{Feynman diagram for the next-to-leading order (NLO) contribution to the grand canonical partition function.}
	\label{Fig:PTDiagramsNLO}
\end{figure}
\begin{figure}[]
	\centering
	\includegraphics[width=120pt]{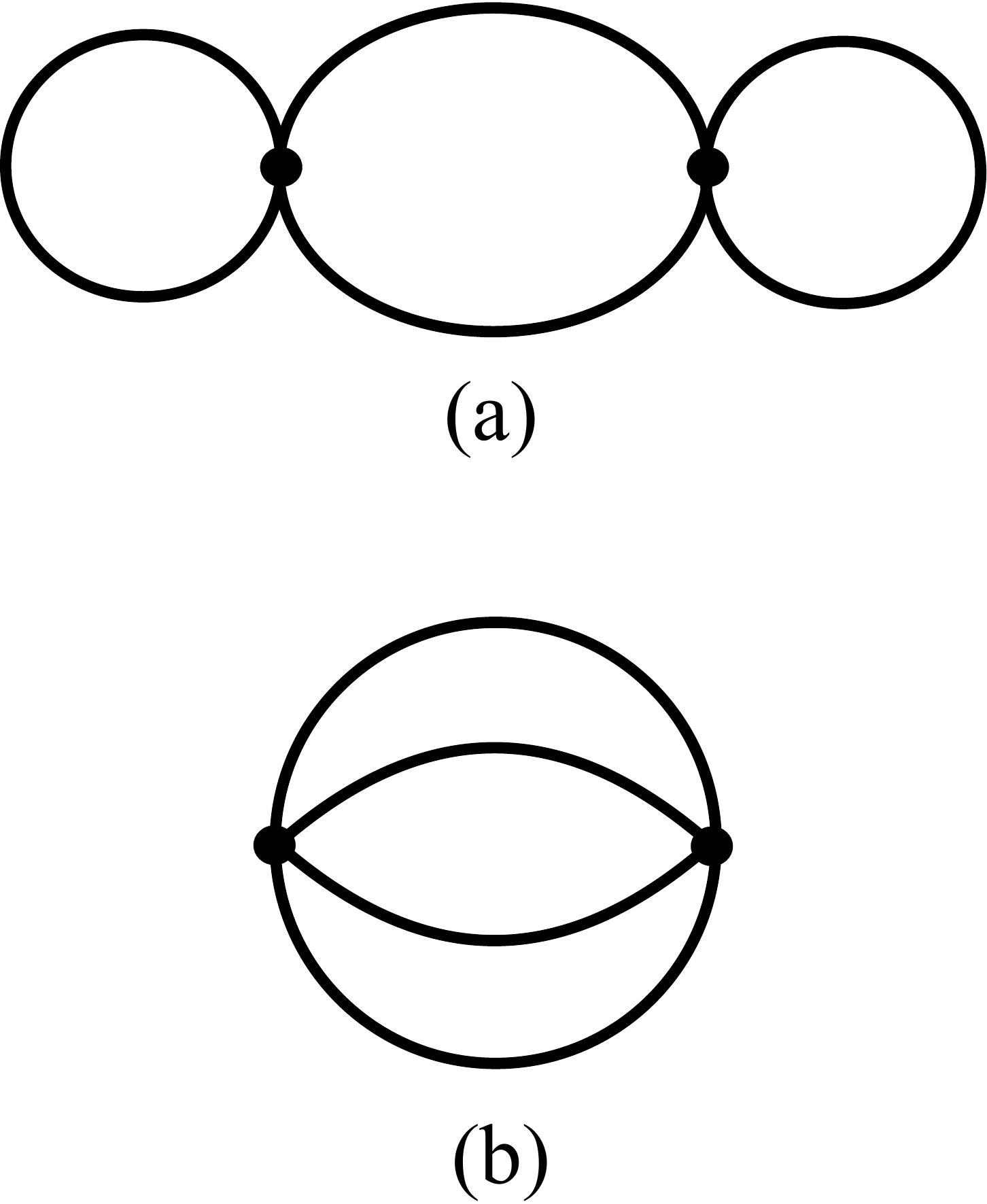}
	\caption{Feynman diagrams for the next-to-next-to-leading order (N2LO) contribution to the grand canonical partition function.}
	\label{Fig:PTDiagramsN2LO}
\end{figure}
\begin{figure}[]
	\centering
	\includegraphics[width=160pt]{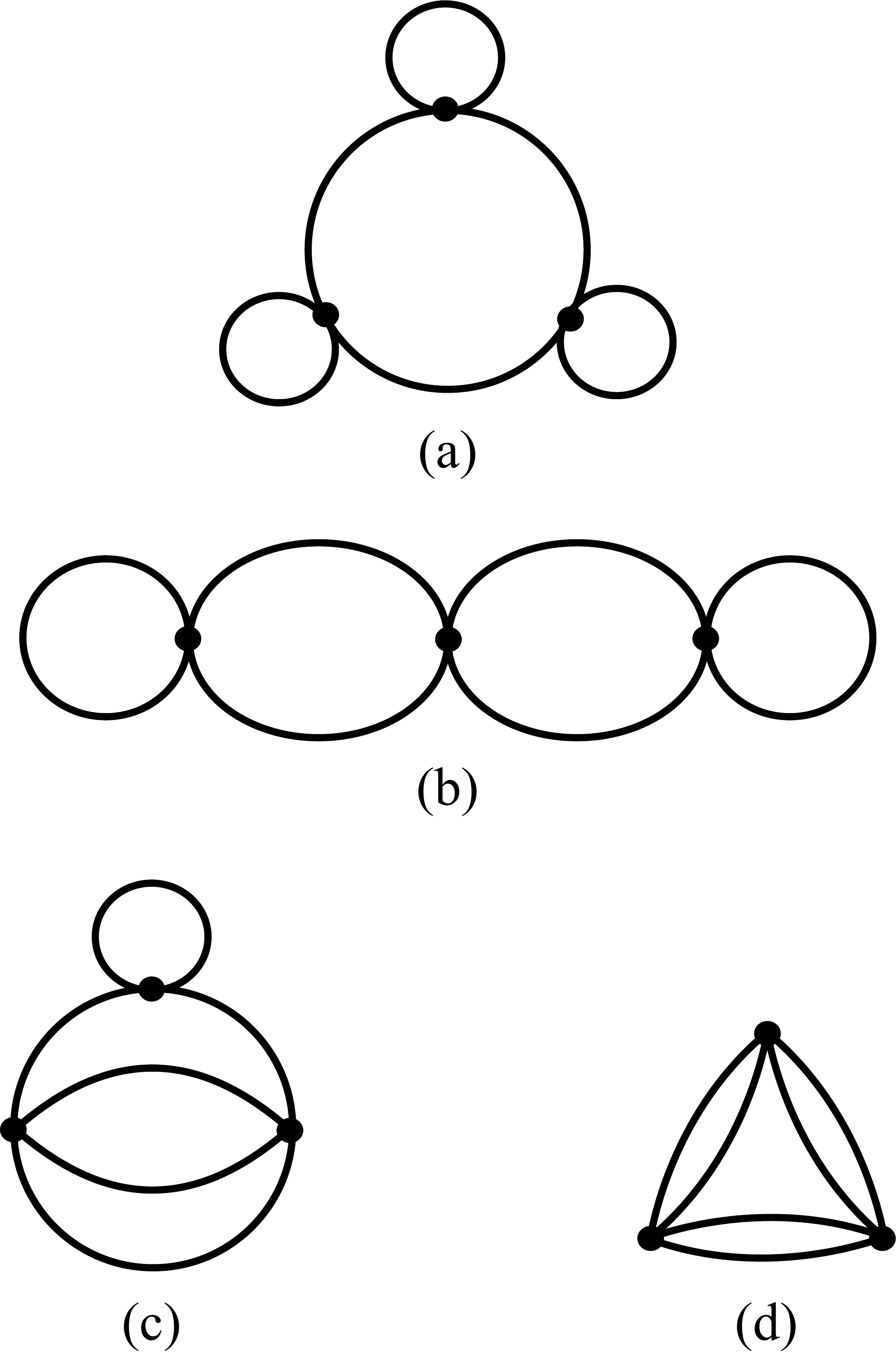}
	\caption{Feynman diagrams for the next-to-next-to-next-to leading order (N3LO) contribution to the grand canonical partition function.}
	\label{Fig:PTDiagramsN3LO}
\end{figure}

Each contribution to the perturbative expansion of the partition function corresponds to a fully connected Feynman diagram of $n$ vertices
(or a product of two or more such diagrams), and can be described by the product of some scalar prefactor with a $1/(N_x N_\tau)^n$ dependence,
and one or more sums over the complete momentum-frequency basis. The Feynman diagrams that appear at NLO, N2LO, and N3LO are provided in Figs. \ref{Fig:PTDiagramsNLO}, \ref{Fig:PTDiagramsN2LO}, \ref{Fig:PTDiagramsN3LO}, respectively.
The analytic expressions for each diagram are provided in Table~\ref{PTContributionTable}, where the Matsubara-frequency sums have already been
carried out using the technique outlined in Sec.~\ref{Section:FrequencySums}; therefore all sums and indices that remain in these expressions are over spatial momenta. The frequency sums always contain $Q_n$ as parameters that encode the spatial momentum dependence:
\beq
Q_n \equiv e^{\tau(\mathbf{k}_n^2/2-\mu)}.
\eeq

A variety of frequency sums for sums over various products of propagators and momentum conservation conditions appear in these expressions. The expressions for $S_1$ and $S_1^{(1)}$ are given earlier in the text by Eqs. (\ref{Eq:S1_0}) and (\ref{Eq:S1_1}), respectively, $S_4$ is derived in Appendix \ref{AppendixBeachBall}, and all others are given in Appendix \ref{Appendix:AllOtherFrequencySums}.

The number of loops, or the number of nested momentum sums that appear for each expression, is also provided explicitly in Table \ref{PTContributionTable}. Since the momentum sums must be computed in full, this number provides an estimate for the computational scaling that is required to compute the value of each diagram; the number of terms that must be computed scales as $N_x^{\ell d}$, where $\ell$ is the number of loops and $d$ is the spatial dimension.

\begin{widetext}
\begin{table*}[t]
\begin{center}
\caption{\label{PTContributionTable} Detail of next-to-leading order (NLO), next-to-next-to-leading order (N2LO), and next-to-next-to-next-to-leading order (N3LO) contributions
(respectively, order $A^2$, $A^4$, and $A^6$)
to the grand-canonical partition function $\mathcal{Z}$. The indicated diagram figure refers to the corresponding fully-connected Feynman diagram or product of such diagrams
for each contribution. The volume $V = N_x N_\tau$ that appears in all expressions refers to the spacetime volume of the lattice. The greatest number of loops that appear for
each fully-connected diagram is also provided. It is implicit in the notation that a momentum-conserving Kronecker delta has been
utilized to eliminate momentum sums: one in the case of $S_4$ and $S_4^{(1)}$, but two in the case of $S_6$.}
\begin{tabular*}{\textwidth}{ @{\extracolsep{\fill}} c c c c c }
\hline
\hline
Order & Diagram Figure & Prefactor & Diagram Frequency Sum & Loops \\
\hline\noalign{\smallskip}
NLO & Fig.~\ref{Fig:PTDiagramsNLO} & $1/(2 \,V)$ & $\left[ \displaystyle\sum\limits_{{\bf p}_1} S_1(Q_1) \right]^2$ & 1-loop \\ \noalign{\smallskip}
\hline\noalign{\smallskip}
N2LO & $[\mathrm{Fig.~\ref{Fig:PTDiagramsNLO}}]^2$ &  $1/(8 \, V^2)$ & $\left[\displaystyle\sum\limits_{{\bf p}_1} S_1(Q_1)\right]^4$ & 1-loop\\ \noalign{\smallskip}
N2LO & Fig.~\ref{Fig:PTDiagramsN2LO}(a) & $-1/(4 \,V^2)$ & $\left[\displaystyle\sum\limits_{{\bf p}_1} S_1^{(1)} (Q_1)\right]\left[\displaystyle\sum\limits_{{\bf p}_1} S_1(Q_1)\right]^2$ & 1-loop \\ \noalign{\smallskip}
N2LO & Fig.~\ref{Fig:PTDiagramsN2LO}(b) & $1/(8\,V^2)$ & $\displaystyle\sum\limits_{{\bf p}_1,{\bf p}_2,{\bf p}_3}S_4(Q_1, Q_2, Q_3)$ & 3-loop \\ \noalign{\smallskip}
\hline\noalign{\smallskip}
N3LO & $[\mathrm{Fig.~\ref{Fig:PTDiagramsNLO}}]^3$ & $1/(48\,V^3)$ & $\left[\displaystyle\sum\limits_{{\bf p}_1} S_1(Q_1)\right]^6$ & 1-loop\\ \noalign{\smallskip}
N3LO & $\mathrm{Fig.~\ref{Fig:PTDiagramsN2LO}(a)}\times \mathrm{Fig.~\ref{Fig:PTDiagramsNLO}}$ & $-1/(8\,V^3)$ & $ \left[\displaystyle\sum\limits_{{\bf p}_1} S_1^{(1)}(Q_1)\right]\left[\displaystyle\sum\limits_{{\bf p}_1} S_1(Q_1)\right]^4$ & 1-loop\\ \noalign{\smallskip}
N3LO & $\mathrm{Fig.~\ref{Fig:PTDiagramsN2LO}(b)}\times \mathrm{Fig.~\ref{Fig:PTDiagramsNLO}}$ & $1/(16\,V^3)$ & $ \left[\displaystyle\sum\limits_{{\bf p}_1,{\bf p}_2,{\bf p}_3} S_4(Q_1, Q_2, Q_3)\right]\left[\displaystyle\sum\limits_{{\bf p}_1} S_1(Q_1)\right]^2$ & 3-loop\\ \noalign{\smallskip}
N3LO & Fig.~\ref{Fig:PTDiagramsN3LO}(a) & $1/(12\,V^3)$ & $\left[\displaystyle\sum\limits_{{\bf p}_1} S_1^{(2)}(Q_1)\right]\left[\displaystyle\sum\limits_{{\bf p}_1} S_1(Q_1)\right]^3$ & 1-loop \\ \noalign{\smallskip}
N3LO & Fig.~\ref{Fig:PTDiagramsN3LO}(b) & $1/(8\,V^3)$ & $\left[\displaystyle\sum\limits_{{\bf p}_1} S_1^{(1)}(Q_1)\right]^2 \left[\displaystyle\sum\limits_{{\bf p}_1} S_1(Q_1)\right]^2$ & 1-loop\\ \noalign{\smallskip}
N3LO & Fig.~\ref{Fig:PTDiagramsN3LO}(c) & $-1/(4\,V^3)$ & $\left[\displaystyle\sum\limits_{{\bf p}_1,{\bf p}_2,{\bf p}_3} S_4^{(1)}(Q_1, Q_2, Q_3)\right]
\left[\displaystyle\sum\limits_{{\bf p}_1} S_1(Q_1)\right]$ & 3-loop\\ \noalign{\smallskip}
N3LO & Fig.~\ref{Fig:PTDiagramsN3LO}(d) & $1/(12\,V^3)$ & $\displaystyle\sum\limits_{{\bf p}_1,{\bf p}_2,{\bf p}_3,{\bf p}_4} S_6(Q_1, Q_2, Q_3, Q_4)$ & 4-loop\\ \noalign{\smallskip}
\hline
\hline
\end{tabular*}
\end{center}
\end{table*}
\end{widetext}

\subsection{Pressure equation of state via perturbation theory}
To determine the pressure $P$, we use the perturbative expansion of the interacting partition function $\mathcal{Z}$, which we expand to order $2n$ in the parameter $A$
and write as
\beq
\mathcal{Z} = \mathcal{Z}_0(1 + A^2\Delta_1 + A^4\Delta_2 + A^6\Delta_3 + \cdots) = e^{\beta P V},
\eeq
where $\Delta_n$ is the N$n$LO contribution, as given by the expressions in Table~\ref{PTContributionTable},
and where $\mathcal{Z}_0$ is the non-interacting grand-canonical partition function,
\beq
\ln \mathcal{Z}_0 = 2 \sum_{k=-N_x/2}^{N_x/2} \ln\left(1 + ze^{-\beta \epsilon_k}\right) = \beta P_0 V,
\eeq
with $\epsilon_k = (2\pi k/N_x)^2 / 2 $.

Thus, the interacting pressure $P$, relative to the non-interacting result $P_0$, becomes
\beq
\frac{P}{P_0} = 1 + \frac{\ln(1 + A^2\Delta_1 + A^4\Delta_2 + A^6\Delta_3 + \cdots)}{\ln\mathcal{Z}_0}.
\eeq
%

To keep the computed value of $\ln\mathcal{Z}$ consistent with the highest order of $A$ in expansion of $\mathcal{Z}$, we expand the numerator $\ln(1+\sum_n A^{2n}\Delta_n)$
in a Taylor series about $A = 0$, such that the expanded form up to N3LO is
\beq
\label{Eq:FinalPTP}
\frac{P}{P_0} = 1 + \frac{1}{\ln\mathcal{Z}_0}\left( A^2\zeta_1 + A^4\zeta_2 + A^6\zeta_3 \right)
\eeq
where
\bea
\zeta_1 &=&  \Delta_1\\
\zeta_2 &=&  \Delta_2 - \frac{1}{2} \Delta_1^2\\
\zeta_3 &=&  \Delta_3 -  \Delta_1 \Delta_2 + \frac{1}{3}\Delta_1^3.
\eea

\begin{figure}[t]
	\centering
	\includegraphics[width=\columnwidth]{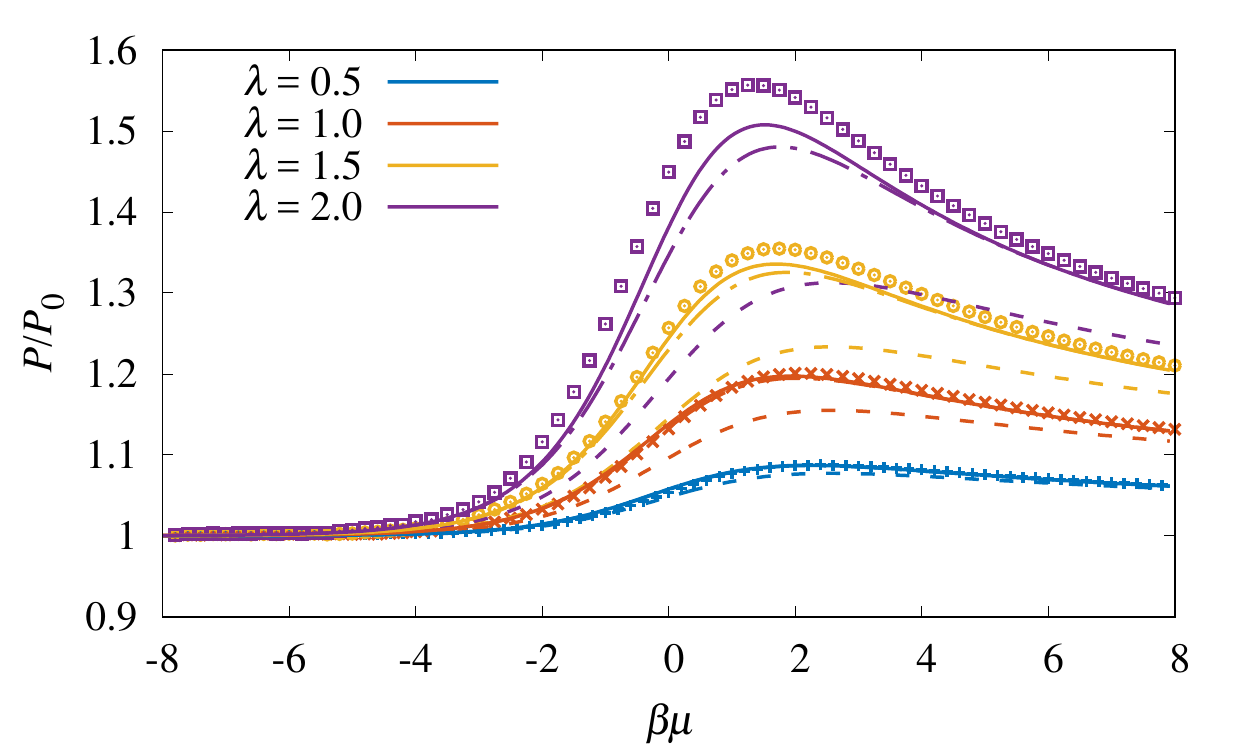}
	\includegraphics[width=\columnwidth]{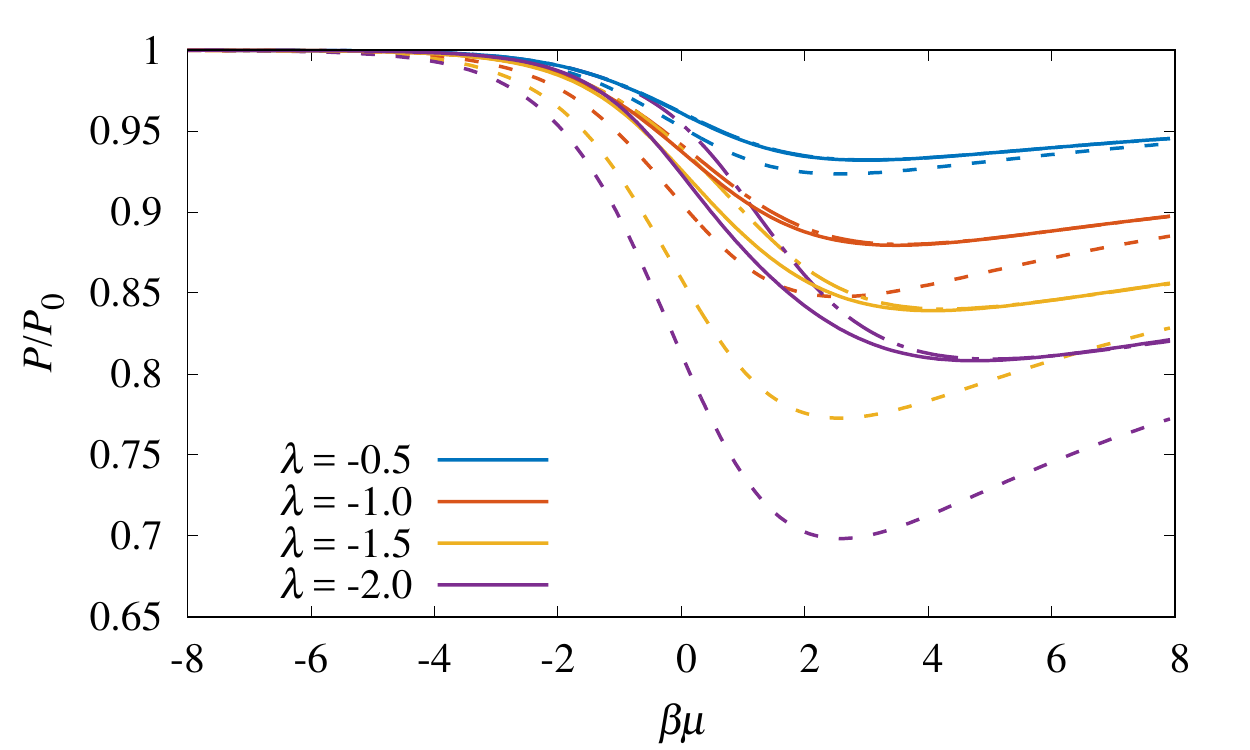}
	\caption{\label{Fig:PTPressure}(Color online) Pressure $P$ of the attractive (top) and repulsive (bottom) unpolarized Fermi gas in units of the pressure of the
	noninteracting system $P_0$, as shown for the dimensionless interaction strengths $\lambda = 0.5$, 1.0, 1.5, 2.0, and $\lambda = -0.5$, -1.0, -1.5, -2.0
	for the attractive and repulsive cases, respectively. The NLO (dashed line), N2LO (dash-dotted line), and N3LO (solid line) results from perturbation
	theory are displayed for each coupling. The corresponding data points for each attractive coupling are computed using HMC (see Ref.~\cite{EoS1D}).}
\end{figure}

Note that while this procedure offers consistency in the order of the coupling at all stages of the calculation, in performing this expansion it is important to be
mindful of the validity of choosing to expand about the non-interacting limit. Since the partition function is an extensive quantity, this expansion may yield
unphysical results in cases where $\sum_n A^{2n}\Delta_n \gg 1$, which may occur in cases where stronger effective interactions are present, as is the case when $N_f > 2$.
However, in all cases shown here, the observables demonstrate physical behavior and are in agreement with alternative methods where available.

Using Eq.~(\ref{Eq:FinalPTP}), we computed the pressure at NLO, N2LO, and N3LO, as a function of the dimensionless parameter $\beta \mu = \ln z$.
Our results for this quantity, for the case of attractive interactions, are shown for a variety of interaction strengths in Fig.~\ref{Fig:PTPressure} (top).
Remarkably, we see evidence of convergence for $\beta \mu \geq 2$, even for $\lambda = 2$. For $\beta \mu \leq 2$, on the other hand, our results
are qualitatively correct but fail to match the Monte Carlo answers by roughly $10\%$ in the worst case of $\lambda = 2$.

A possible explanation of this behavior is that perturbation theory simply fails to capture the effect of the two-body bound state on the virial coefficients
as $\lambda$ is increased, which in the case of $b_2$ involves a dominant inverse gaussian contribution $\propto \exp(\lambda^2/4)$. This effect could
be absorbed into the definition of $\lambda$ by way of a new renormalization scheme defined by matching to an exact calculation of $b_2$ (lattice or continuum),
instead of identifying the coupling $g$ with the inverse scattering length.
Clearly, such a renormalization scheme would improve the agreement with the Monte Carlo results in the semiclassical region of negative $\beta\mu$, but by
the same token it would spoil it for $\beta \mu \geq 2$. Nevertheless, as the perturbative expansion is extended beyond third-order, the convergence properties
of the regime is expected to improve.

In all cases, the NLO results are quantitatively disappointing, but N2LO and N3LO display substantial improvement both in convergence
and in approaching the Monte Carlo results. Naturally, this improvement is not free: the computational effort increases dramatically for N2LO and N3LO relative
to NLO. Similar behavior is seen for the repulsive case in the bottom of Fig.~\ref{Fig:PTPressure} but with an important difference: the results oscillate as the
perturbative order is increased, whereas in the attractive case convergence appears to be monotonic.

\subsection{Density equation of state via perturbation theory and complex Langevin}

\begin{figure}
	\centering
	\includegraphics[width=\columnwidth]{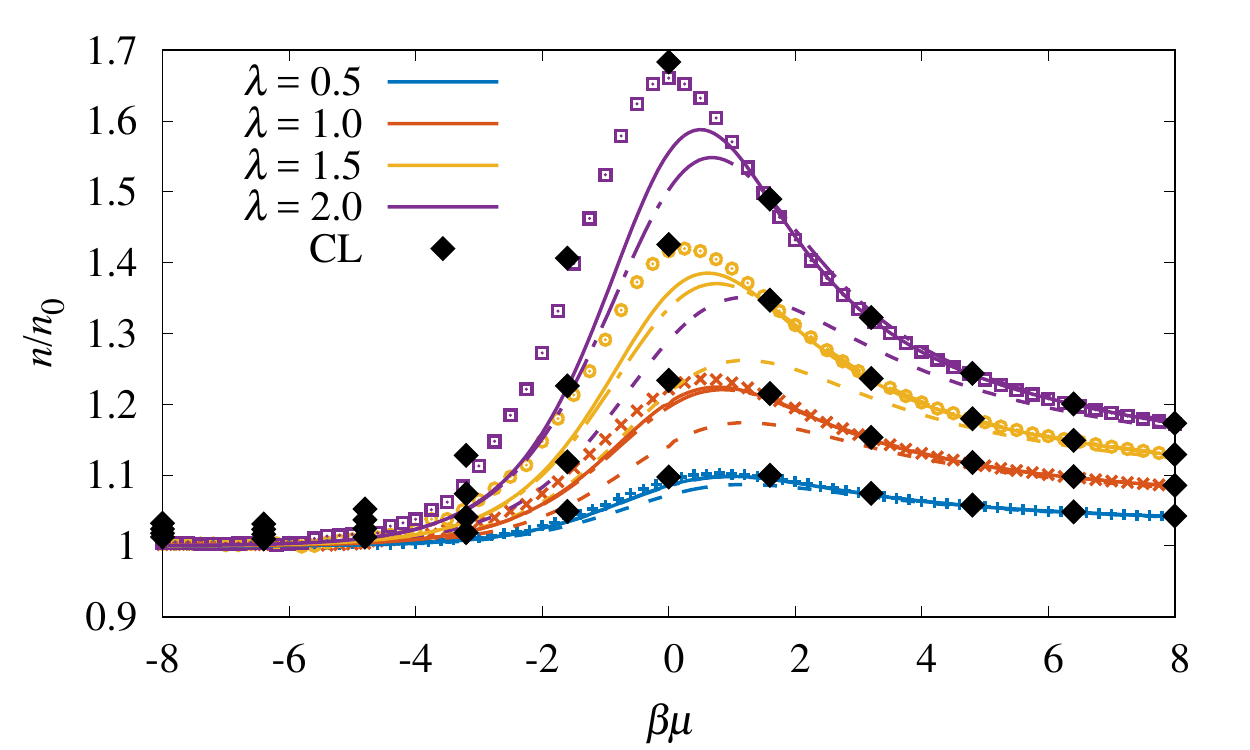}
	\includegraphics[width=\columnwidth]{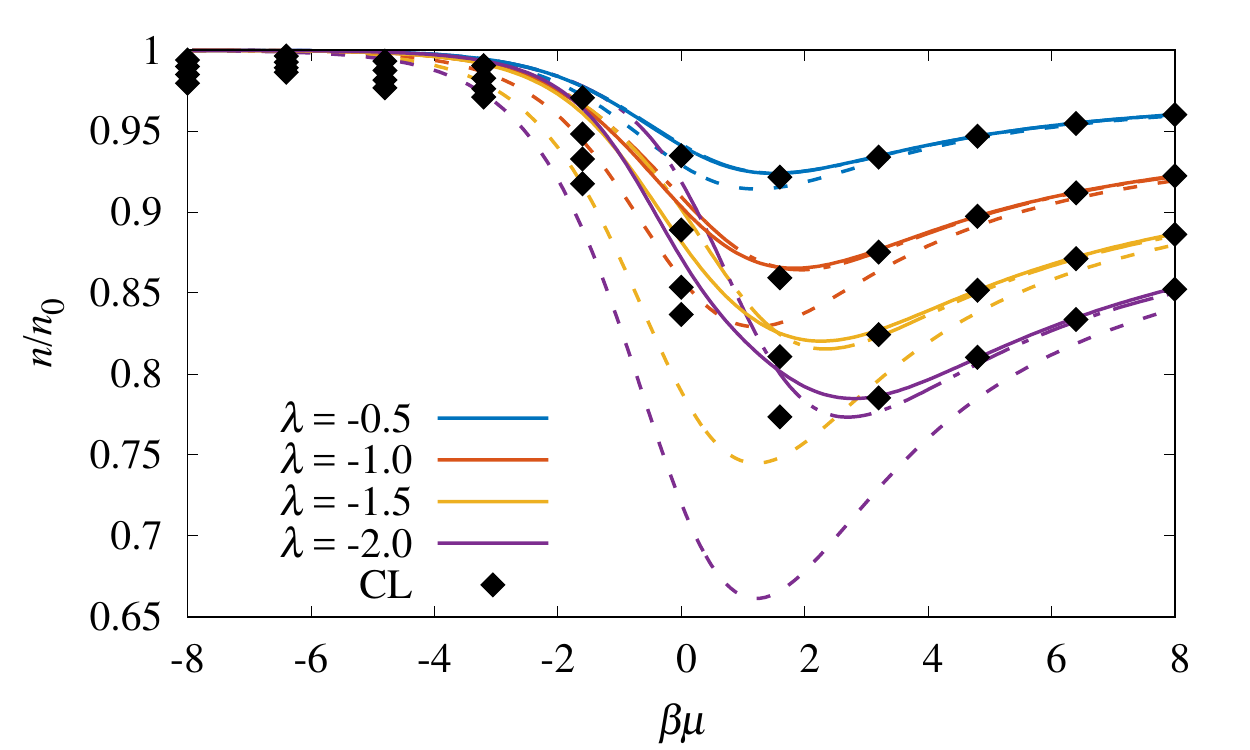}
	\caption{\label{Fig:PTDensity}(Color online) Density $n$ of the attractive (top) and repulsive (bottom) unpolarized Fermi gas in units of the density of the
	noninteracting system $n_0$, as shown for the dimensionless interaction strengths $\lambda = 0.5$, 1.0, 1.5, 2.0 (attractive), and $\lambda = -0.5$, -1.0, -1.5, -2.0
	(repulsive). The NLO (dashed line), N2LO (dash-dotted line), and N3LO (solid line) results of perturbation theory are displayed for each coupling and are
	compared with HMC results (see Ref.~\cite{EoS1D}) in the attractive case. For both plots, the black diamonds show CL results
	(RL for the attractive case), regulated with $\xi=0.1$ as described in the main text. The statistical uncertainty of the CL results is estimated to be on
	the order of the size of the symbols, or less, as supported by the smoothness of those results.}
\end{figure}

Based on our knowledge of the pressure, differentiation with respect to $\beta \mu$ gives access to the density, for which results from perturbation theory
can be compared more directly with quantum Monte Carlo results (the pressure is not typically a quantity computed directly in Monte Carlo calculations,
but is rather obtained by integrating the density; see Ref.~\cite{EoS1D} for details).
We show such a comparison in Fig.~\ref{Fig:PTDensity} (top) for the case of attractive interactions where Monte Carlo calculations are possible without a sign problem.
For repulsive interactions, our calculations yield the results shown in Fig.~\ref{Fig:PTDensity} (bottom). In both cases, we have numerically differentiated
with a point spacing of $d\mu = 0.01$.\\

Both figures also display density results obtained using CL techniques. Remarkably, this method demonstrates
excellent agreement with both hybrid Monte Carlo and perturbation theory for attractive interactions (where CL becomes RL),
and perturbative results for repulsive interactions. The auxiliary field $\sigma$ was evolved with CL dynamics for $10^5$
iterations with an adaptive timestep $\delta t$, adjusted by monitoring the magnitude of the CL drift.

\subsection{Perturbative virial coefficients}

Based on the results shown above, we implement a method of particle-number projection, which is well known
in the area of nuclear physics, to calculate virial coefficients perturbatively. To explain the method, we recall that the virial expansion for the pressure is given by
\beq
-\beta P V = \ln \mathcal Z = \mathcal {Q}_1 \sum_{n = 1}^{\infty} b_n z^n, \label{Eq:VirialExpansion}
\eeq
where $\mathcal {Q}_1 = N_f L / \lambda_T$ is the one-particle canonical partition function (here $\lambda_T= \sqrt{2\pi \beta}$ is the thermal
de Broglie wavelength), and $b_n$ are the virial coefficients we want to extract.

Using our (semi-)analytic expressions for $ \ln \mathcal Z(z) $, it is possible to extract the coefficients $b_n$ via a numerical Fourier projection,
\beq
b_n = \frac{1}{ \mathcal Q_1 \alpha^n } \int_0^{2\pi}\! \frac{d\phi}{2 \pi}  e^{-i n \phi} \ln \mathcal Z(z\rightarrow \alpha e^{i\phi}),
\eeq
where the partition function is evaluated at $z = \alpha\exp(i\phi)$ and $\alpha$ is an arbitrary real parameter that is independent of the result for $b_n$, but is used to ensure that a well-behaved integrand is used. A value of $\alpha = 0.01$ is well-suited for the evaluations performed here.

\begin{figure}
	\centering
	\includegraphics[width=\columnwidth]{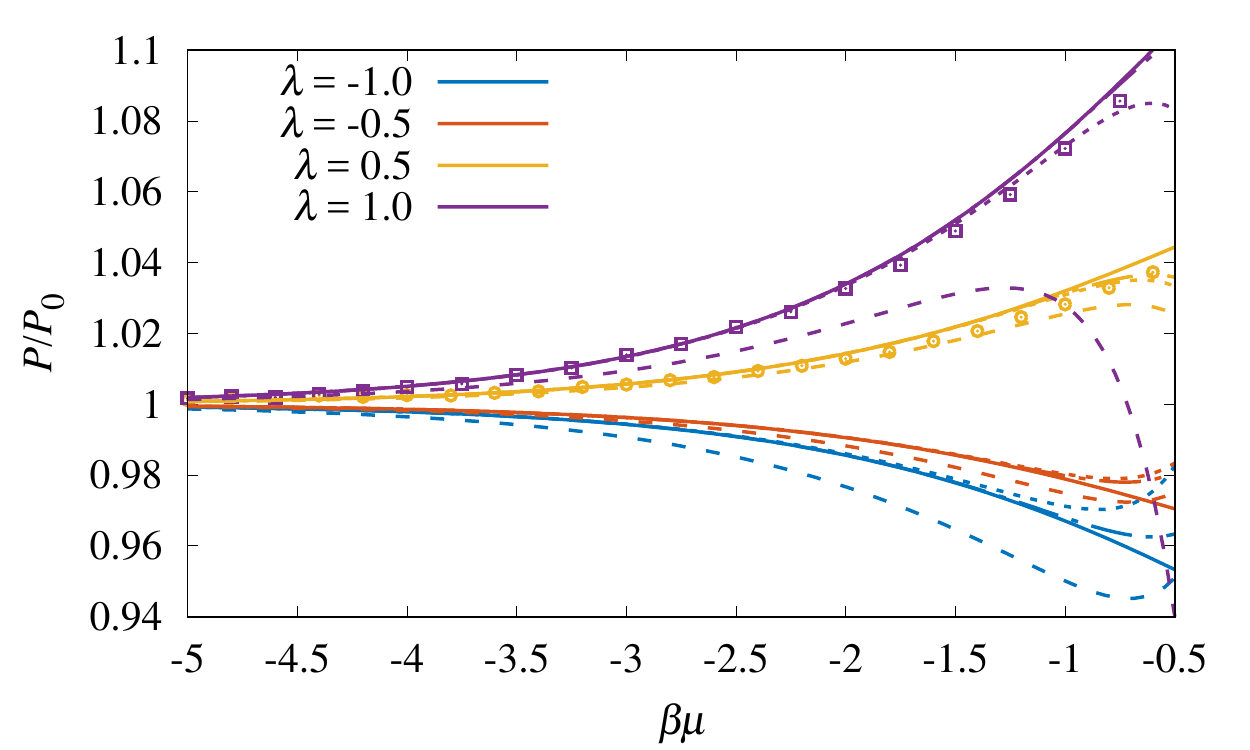}
	\caption{\label{Fig:PTVirialExpansion}(Color online) Fifth-order perturbative virial expansion for the pressure $P/P_0$ for two attractive
  and repulsive couplings. The expansions are shown for which the virial coefficients are computed at NLO (dashed line), N2LO, (dotted line),
  and N3LO (dash-dotted line). The fully solid line shows the full perturbative calculation of the pressure at N3LO, as shown in
  Fig.~\ref{Fig:PTPressure}. Results from HMC data (see Ref.~\cite{EoS1D}) are displayed as corresponding data points for the attractive couplings.}
\end{figure}

\begin{table}[t]
\caption{\label{PTVirialCoefficients} Results for the second, third, fourth, and fifth-order virial coefficients of the pressure
 $P$ [see Eq.~(\ref{Eq:VirialExpansion})] at NLO, N2LO, and N3LO for two repulsive and attractive couplings. All coefficients
 are computed for a spatial lattice size of $N_x = 100$, $\beta = 8.0$, and a temporal lattice spacing of $\tau = 0.05$.
 The fifth-order virial expansion is displayed in Fig. \ref{Fig:PTVirialExpansion}.}
\begin{tabular*}{\columnwidth}{ @{\extracolsep{\fill}} d c d d d }
\hline
\hline
\lambda & Coefficient & \text{NLO} & \text{N2LO} & \text{N3LO} \\
\hline
 \multirow{4}{*}{-1.0}  &  $b_2$ & -0.551 & -0.473 & -0.473 \\
   &  $b_3$  & 0.472 & 0.373 & 0.342 \\
   &  $b_4$  & -0.452 & -0.380 & -0.280\\
   &  $b_5$  & 0.449 & 0.446 & 0.246\\
 \hline
 \multirow{4}{*}{-0.5}  &  $b_2$  & -0.453 & -0.433 & -0.433\\
   &  $b_3$  & 0.333 &  0.308 & 0.304 \\
   &  $b_4$  & -0.289 & -0.271 & -0.259\\
   &  $b_5$  & 0.270 & 0.269 & 0.244\\
 \hline
 \multirow{4}{*}{0.5}  &  $b_2$  & -0.253 & -0.233 & -0.233 \\
   &  $b_3$  & 0.0508 & 0.0254 & 0.0295\\
   &  $b_4$  & 0.0408 & 0.0592 & 0.0461\\
   &  $b_5$  & -0.0925 & -0.0932 & -0.0671\\
 \hline
 \multirow{4}{*}{1.0}  &  $b_2$  & -0.152 & -0.0712 & -0.0712\\
   &  $b_3$  &  -0.0922 & -0.194 & -0.161\\
   &  $b_4$  &  0.208 & 0.282 & 0.176\\
   &  $b_5$  &  -0.276 & -0.279 & -0.0677\\
\hline
\hline
\end{tabular*}
\end{table}

The perturbative expansion of virial coefficients $b_n$ displays signs of convergence at weak coupling $|\lambda| \leq 1$
and for low enough virial order $n$. We find that for $\lambda=\pm1$ the predictions for $b_5$ cease to
converge, at least at N3LO. For comparison, an exact lattice calculation of the second-order virial coefficient at $N_x=100$
and $\beta = 8$ under attractive interactions (see Ref.~\cite{EoS1D}) shows that for $\lambda = 0.5$, $b_2 = -0.230$, and for $\lambda = 1.0$,
$b_2 = -0.0351$. This discrepancy compared with the perturbative value for $b_2$, particularly for larger values of the coupling,
underscores the difficulty in quantitatively capturing the behavior in the virial region using perturbation theory.


\section{Summary and Conclusions~\label{Sec:Conclusions}}

In summary, we have determined, using both up to third order in lattice perturbation theory and CL techniques, the pressure and
density equations of state of non-relativistic spin-$1/2$ fermions with contact interactions in both attractive
and repulsive regimes. We have used those results in combination with particle-number projection techniques
to obtain a perturbative expansion for virial coefficients.

Our results for the perturbative pressure and density display some remarkable features: even for couplings as high as $\lambda=2$
there is a region of fugacity values (namely for $\beta \mu \geq 1$) that displays clear signs of convergence
toward excellent agreement with extant non-perturbative Monte Carlo results (for the case of attractive interactions).

We performed our perturbative lattice calculations by following an unconventional route
based on first decoupling the interaction via a Hubbard-Stratonovich transformation, and then undoing that
transformation {\it after} expanding the non-interacting fermion determinant under the path integral.
The approach is interesting because it is amenable to automation in a way that
seems easier to handle than conventional approaches that use operators throughout.

Additionally, we have presented a way to carry out the Matsubara-frequency sums that is both simple and efficient,
in particular in that it does not involve contour integration in the complex plane and in that it treats all the nested sums
simultaneously. While techniques that are similar in spirit exist in the continuum, we are not aware of this
type of approach for systems on the lattice. Carrying out those frequency sums is more than a matter of convenience: it is
essential in order to obtain numerically manageable expressions, namely expressions that only contain spatial
momentum sums. Furthermore, we have shown how a common trick, namely differentiation with respect to an
auxiliary parameter, can generate expressions needed for certain diagrams using expressions for lower-order diagrams.
The frequency sums presented above and in the appendices are universal in that they apply to systems regardless
of the dispersion relation, external potential, and interaction potential, as long as the latter is energy-independent.

To access repulsive interactions in a non-perturbative fashion (a well-known case suffering from the sign problem), we
implemented the CL method and proposed a modification to the action that prevents uncontrolled excursions into the
complex plane. This modified action yields a damping term in the CL dynamics which, in turn, introduces
systematic effects of a priori unknown size. Our investigations of those effects, both on the attractive and repulsive sides, support the idea that
the impact on the density equation of state is within the uncertainties of our calculations; other observables would require
independent investigations. The effectiveness of this approach in other situations, like polarized matter and higher dimensions,
remains to be studied; however, nothing seems to prevent us from using the same idea in those cases.
For the 1D system studied here, the CL results agree with the perturbative ones in regions where perturbation
theory appears to converge (within the orders studied here); everywhere else, the CL answers represent non-perturbative predictions.

\acknowledgments
We thank G. Aarts, J. Braun, and A. Vuorinen for comments on the manuscript.
This material is based upon work supported by the
National Science Foundation under Grants No.
DGE{1144081} (Graduate Research Fellowship Program),
PHY{1306520} (Nuclear Theory Program), and
PHY{1452635} (Computational Physics Program).

\appendix
\begin{widetext}
\section{\label{AppendixBeachBall} Matsubara frequency sum derivation example of $S_4$}
To further illustrate our method of analytically evaluating Matsubara frequency sums on the lattice, we will show in detail how to calculate the case of $S_4$. Note that higher-order frequency sums, such as $S_4^{(1)}$, can also be determined from the expressions derived here by inserting a parameter $\lambda$ such that, for example, $Q_1 \rightarrow Q_1/\lambda$, and proceed to calculate derivatives with respect to $\lambda$, evaluated at $\lambda = 1$. This is a generalization of the procedure used for the case of $S_1^{(1)}$, as illustrated in Eq.~(\ref{Eq:S1LambdaInsertion}).\\

The frequency sum $S_4$ is defined as the sum over four free propagators, where the indices of the first three propagators are free, and the fourth is constrained by the momentum conservation condition $\omega_4 = \omega_1 - \omega_2 + \omega_3$. Therefore, we write the expression for $S_4$ with a Kronecker delta such that
\bea
S_4 = \!\!\!
\sum_{\{n\}=1}^{N_\tau}
\left[
\prod_{q=1}^{4}
\frac{1}{1 - Q_q e^{i \omega_{n_q}}}
\right]
\delta(\omega_{n_3} \!-\! \omega_{n_2} \!+\! \omega_{n_1} \!-\! \omega_{n_4}).
\eea

We first represent the Kronecker delta with a sum over the complete frequency-space basis and expand the denominators using a geometric series,
\bea
S_4 = \frac{1}{N_\tau}
\sum_{p=0}^{N_\tau-1}
\!\!
\sum_{\{n\}=1}^{N_\tau}
\!\!
\left[
\prod_{q=1}^{4}
\sum_{k_q = 1}^{\infty} Q^{k_q}_q e^{i k_q \omega_{n_q}} \!\!
\right]\!\!
e^{i p (\omega_{n_3} - \omega_{n_2} + \omega_{n_1} - \omega_{n_4})}.
\eea

Next, we reorder the sums in preparation to carry out all the frequency sums, which
result in delta functions. Note that in the process, the appearance of powers of $(-1)$, which
results from the fermionic Matsubara frequencies.

\bea
S_4
&=&
\frac{1}{N_\tau}
\sum_{p=0}^{N_\tau-1}
\sum_{\{k\}=0}^{\infty} Q^{k_1}_1 Q^{k_2}_2 Q^{k_3}_3 Q^{k_4}_4
\sum_{\{n\}=1}^{N_\tau}
e^{i \omega_{n_1}(k_1 + p)}
e^{i \omega_{n_2}(k_2 - p)}
e^{i \omega_{n_3}(k_3 + p)}
e^{i \omega_{n_4}(k_4 - p)}
\\
&=&
N_\tau^3
\sum_{p=0}^{N_\tau-1}
\sum_{\{m\}=-\infty}^{\infty}
\sum_{\{k\}=0}^{\infty} Q^{k_1}_1 Q^{k_2}_2 Q^{k_3}_3 Q^{k_4}_4 (-1)^{m_1+m_2+m_3+m_4}
\times \\
&& \nonumber
\delta(k_1 + p - m_1 N_\tau)
\delta(k_2 - p - m_2 N_\tau)
\delta(k_3 + p - m_3 N_\tau)
\delta(k_4 - p - m_4 N_\tau)
\eea

To saturate the delta functions when summing over $\{ k\}$, we must account for the fact that all $k$ are positive.
To this end, we extend the sums to negative values of $k$ and insert Heaviside functions accordingly (defined to saturate
for non-negative values of the argument), which results in
\bea
S_4&=&
N_\tau^3
\sum_{p=0}^{N_\tau-1}
\sum_{\{m\}=-\infty}^{\infty}
Q^{N_\tau m_1}_1 Q^{N_\tau m_2}_2 Q^{N_\tau m_3}_3 Q^{N_\tau m_4}_4
(-1)^{m_1+m_2+m_3+m_4}
\left(\frac{Q^{}_2 Q^{}_4}{Q^{}_1 Q^{}_3}\right)^p
\times \\
&& \nonumber
\theta(- p + m_1 N_\tau)
\theta(p + m_2 N_\tau)
\theta(- p + m_3 N_\tau)
\theta(p + m_4 N_\tau)
\eea

Because of the range of $p$, the $\theta$ functions imply the
following conditions: $m_1 \geq 0$ if $p=0$, but otherwise $m_1 > 0$; an identical condition for $m_3$; $m_2 \geq 0$; and $m_4 \geq 0$.
Using the last two conditions, we obtain a first simplification:

\bea
S_4 =
N_\tau^3
\frac{1}{1 + Q^{N_\tau}_2}
\frac{1}{1 + Q^{N_\tau}_4}
\sum_{p=0}^{N_\tau-1}
\sum_{\{m\}=-\infty}^{\infty}
Q^{N_\tau m_1}_1  Q^{N_\tau m_3}_3
(-1)^{m_1+m_3}
\left(\frac{Q^{}_2 Q^{}_4}{Q^{}_1 Q^{}_3}\right)^p
\theta(- p + m_1 N_\tau)
\theta(- p + m_3 N_\tau)
\eea

Applying the remaining two conditions amounts to (note we separate the $p=0$ and $p\neq 0$ cases in the second line)
\bea
&& \!\!\!\!\!\!\!\!
\sum_{\{m\}=-\infty}^{\infty}
Q^{N_\tau m_1}_1  Q^{N_\tau m_3}_3
(-1)^{m_1+m_3}
\sum_{p=0}^{N_\tau-1}X^p
\theta(-p \!+\! m_1 N_\tau)
\theta(-p\!+\! m_3 N_\tau)
\\
&=&
\frac{1}{1 + Q^{N_\tau}_1}
\frac{1}{1 + Q^{N_\tau}_3}
\left [
1 +
{Q^{N_\tau}_1}
{Q^{N_\tau}_3}
\sum_{p=1}^{N_\tau-1}
X^p
\right] \nonumber
\\
&=&
\frac{1}{1 + Q^{N_\tau}_1}
\frac{1}{1 + Q^{N_\tau}_3}
\left [
1 +
{Q^{N_\tau}_1}
{Q^{N_\tau}_3}
\left(
\frac{1\!-\! X^{N_\tau}}{1 \!-\! X}\!-\! 1
\right)
\right] \nonumber
,
\eea

where $X = {Q^{}_2 Q^{}_4}/({Q^{}_1 Q^{}_3})$.
Thus, the final result is simply
\bea
S_4 =
N_\tau^3
\left(\prod_{k=1}^4
\frac{1}{1 + Q^{N_\tau}_k}\right)
\left [
1 +
({Q_1}{Q_3})^{N_\tau}
\left(
\frac{1\!-\! X^{N_\tau}}{1 \!-\! X}\!-\! 1
\right)
\right].
\eea

Note that, as anticipated, we obtained this result without resorting to contour integration, and moreover we addressed all the frequency sums simultaneously.

\section{Listing of Higher-Order Matsubara Frequency Sums}
\label{Appendix:AllOtherFrequencySums}
In this section, we will provide a listing of all Matsubara frequency sums that appear in the perturbative contributions to the grand-canonical partition function at third-order.
Note that the first- and second-order sums $S_1$, $S_1^{(1)}$, and $S_4$ are provided elsewhere in the text.
In all cases, our expressions were checked by comparing with a direct evaluation of the defining sums in a small spacetime volume.

The frequency sum $S_1^{(2)}$ is defined as the sum over the product of three identical propagators, and the derived expression is given by
\beq
S_1^{(2)} = \sum_{n=1}^{N_\tau} \left(\frac{1}{1-Q_1 e^{i\omega_n}}\right)^3 = \frac{N_\tau (2-Q_1^{N_\tau}[N_\tau^2 + 3N_\tau - 4] + Q_1^{2N_\tau}[N_\tau^2 - 3 N_\tau + 2])}{2(1+Q_1^{N_\tau})^3}.
\eeq
The frequency sum $S_4^{(1)}$ can be obtained, as explained in the main text, by using a differentiation trick:
\beq
S_4^{(1)} =
- \left.\frac{d S_4(\lambda)}{d\lambda}\right |_{\lambda=1}
\eeq
where $S_4^{}(\lambda) \equiv \lambda^{-1} \left. S_4^{} \right|_{Q_1 \to Q_1/\lambda}$. Thus,

\begin{multline}
S_4^{(1)}=
\sum_{n_1,n_2,n_3,n_4=1}^{N_\tau} \left(\frac{1}{1-Q_1 e^{i\omega_{n_1}}}\right)^2 \frac{1}{1-Q_2 e^{i\omega_{n_2}}} \frac{1}{1-Q_3 e^{i\omega_{n_3}}} \frac{1}{1-Q_4 e^{i\omega_{n_4}}}\delta(\omega_{n_1}-\omega_{n_2}+\omega_{n_3}+\omega_{n_4})\\
= N_\tau^3\left(\prod_{k = 1}^4 \frac{1}{1+Q_k^{N_\tau}} \right)
\left\{
\left[1+(Q_1 Q_3)^{N_\tau}\left(\frac{1-X^{N_\tau}}{1-X} - 1\right)\right]
\left(\frac{(N_\tau-1) Q_1^{N_\tau} - 1}{1+Q_1^{N_\tau}}\right)\right . \\
\left . + N_\tau (Q_1 Q_3)^{N_\tau}\left[ 1 - \frac{1-X^{N_\tau}}{1-X} - \frac{X^{N_\tau}}{1-X} + \frac{X}{N_\tau}\frac{1-X^{N_\tau}}{(1-X)^2} \right] \right \}
\end{multline}
where $X \equiv Q_2 Q_4 / (Q_1 Q_3)$.

Finally, the frequency sum $S_6$ is defined as
\begin{multline}
S_6 = \sum_{\substack{n_1,n_2,n_3,\\ n_4,n_5,n_6}}\left(\prod_{k = 1}^6\frac{1}{1-Q_k e^{i\omega_{n_k}}}\right)\delta(\omega_{n_1}-\omega_{n_2}-\omega_{n_5}+\omega_{n_6}) \, \delta(\omega_{n_3}-\omega_{n_4}+\omega_{n_1}-\omega_{n_6})\\
= N_\tau^4 \left(\prod_{k = 1}^6 \frac{1}{1+Q_k^{N_\tau}}\right)\left(S_{6A} + S_{6B} + S_{6C} + 1\right)
\end{multline}
such that
\beq
S_{6A} \equiv (Q_2 Q_4)^{N_\tau} \left[\left(\frac{1-(XY)^{N_\tau}}{1-XY}-1\right)(1+Q_5^{N_\tau}+Q_6^{N_\tau})+F(X,Y)Q_6^{N_\tau}+F(Y,X)Q_5^{N_\tau}\right],
\eeq
\beq
S_{6B} \equiv (Q_4 Q_6)^{N_\tau}\left(\frac{1-Y^{N_\tau}}{1-Y}-1\right), \quad \mathrm{and} \quad S_{6C} \equiv (Q_2 Q_5)^{N_\tau}\left(\frac{1-X^{N_\tau}}{1-X}-1\right),
\eeq
where we also define
\beq
F(X,Y) \equiv \frac{1-Y^{N_\tau}}{1-Y}\left(1-\frac{1-X^{N_\tau}}{1-X}\right) + \frac{1}{1-Y}\left(\frac{1-X^{N_\tau}}{1-X}-\frac{1-(XY)^{N_\tau}}{1-XY}\right),
\eeq
as well as $X \equiv Q_1 Q_6 / (Q_2 Q_5)$ and $Y \equiv Q_3 Q_5 / (Q_4 Q_6)$.
As evident from the above expressions, it is important to consider limiting cases near poles (e.g. $X=1$) when numerically evaluating these sums;
such limits are straightforward to compute and implement.

\end{widetext}


\end{document}